\documentclass[english]{article}
\usepackage[T1]{fontenc}
\usepackage[latin9]{inputenc}
\usepackage{mathrsfs}
\usepackage{amsmath}
\usepackage{cancel}
\usepackage{esint}

\makeatletter
\usepackage{fullpage}

\makeatother

\usepackage{babel}
\begin{document}
\title{Internal symmetry in Poincarè gauge gravity}
\author{James T. Wheeler\thanks{Utah State University Department of Physics, Logan, UT, USA, jim.wheeler@usu.edu,
orcid 0000-0001-9246-0079}}
\maketitle
\begin{abstract}
We find a large internal symmetry within 4-dimensional Poincarè gauge
theory.

In the Riemann-Cartan geometry of Poincaré gauge theory the field
equation and geodesics are invariant under projective transformation,
just as in affine geometry. However, in the Riemann-Cartan case the
torsion and nonmetricity tensors change. By generalizing the Riemann-Cartan
geometry to allow both torsion and nonmetricity while maintaining
local Lorentz symmetry the difference of the antisymmetric part of
the nonmetricity Q and the torsion T is a projectively invariant linear
combination S = T - Q with the same symmetry as torsion. The structure
equations may be written entirely in terms of S and the corresponding
Riemann-Cartan curvature. The new description of the geometry has
manifest projective and Lorentz symmetries, and vanishing nonmetricity.

Torsion, S and Q lie in the vector space of vector-valued 2-forms.
Within the extended geometry we define rotations with axis in the
direction of $S$. These rotate both torsion and nonmetricity while
leaving $S$ invariant. In n dimensions and (p, q) signature this
gives a large internal symmetry. The four dimensional case acquires
SO(11,9) or Spin(11,9) internal symmetry, sufficient for the Standard
Model.

The most general action up to linearity in second derivatives of the
solder form includes combinations quadratic in torsion and nonmetricity,
torsion-nonmetricity couplings, and the Einstein-Hilbert action. Imposing
projective invariance reduces this to dependence on S and curvature
alone. The new internal symmetry decouples from gravity in agreement
with the Coleman-Mandula theorem. 

\pagebreak{}
\end{abstract}

\section{Introduction}

Geodesics are the preferred paths followed by test particles in general
relativity and much of what we know about gravity comes from analyzing
these paths. Although relativity gives us proper time as a preferred
parameter, even within special relativity we may use any observer's
time instead. It is therefore natural to examine the effect of reparameterizing
geodesics. The resulting projective transformation \cite{Thomas}
of the connection is a well-known symmetry of the curvature of general
relativity. However, even in the most restricted form of general relativity,
the relationship between projective symmetry and the metric is nontrivial.
Fully confronting the conflict between the preferred proper time given
by the metric and the obvious freedom to reparameterize suggests the
desirability of a reparameterization invariant form of general relativity.

Such compatibility was achieved in a widely known work by Ehlers,
Pirani, and Schild \cite{EPS} and honed in a study by Matveev and
Trautman \cite{Matveev1} and later Matveev and Scholz \cite{Matveev2}.
In these papers it is argued that we determine the geometry of spacetime
by studying the geodesics of timelike and null geodesics. Concretely,
these authors show that we can infer a projective connection from
a knowledge of the timelike geodesics of test particles, while the
same program for light following null curves determines a conformal
connection. Agreement between the two connections in the limit of
high velocities leads to an integrable Weyl geometry \cite{Wheeler2018},
that is, a geometry with dilatational symmetry which becomes Riemannian
with a particular choice of units for proper time.

Given this satisfactory and minimal resolution, we are free to specify
the Riemannian gauge and carry out gravitational studies as usual.
However, the development of general relativity as a Poincarè gauge
theory over the last two-thirds of a century opens some new possibilites.
It is these alternative possibilities that we examine here. We explore
the combination of projective and Lorentz symmetries starting from
a Cartan formulation of gravity. The possibilities include the integrable
Weyl form of general relativity, and the integrable Weyl form of the
Einstein, Cartan, Sciama, Kibble (ECSK) generalization. The ECSK theory
includes field equations driven by both mass and spin. But working
in the newer formalism suggests a further, deeper symmetry.

By freeing the connection from the metric, we are led to consider
two new fields.

Within Poincarè gauge theory it is natural to include torsion as well
as curvature. When fermionic matter is present the torsion becomes
the geometric equivalent of spin density in the same way that the
Einstein tensor is the geometric equivalent of energy. There is a
pleasing justice to this because mass and spin are the Casimir invariants
of the Poincarè Lie algebra. But the observation is puzzling because
the experimental limits on torsion are strong \cite{Neville 1980,Shapiro,BelyaevShapiro}.
This leads to much of the research on Riemann-Cartan geometries being
devoted to understanding why torsion effects should be absent or negligible.

The second new field, the nonmetricity, reflects the compatibility
of the metric and connection. Since Poincarè gauge theory naturally
makes the metric and connection independent, and because the integrable
Weyl geometry found in \cite{Matveev1,Matveev2} gives the nonmetricity
a nonvanishing but removable trace, it is sensible to consider a formulation
of gravity in which nonmetricity is free to play a role.

Our goal is to understand the context of general relativity while
preserving its overwhelming success as the formulation of gravity.
By allowing torsion and nonmetricity within Poincarè gauge theory,
we find a surprisingly large internal symmetry. This new symmetry
is present even when the torsion and nonmetricity vanish as long as
we admit them as possible within the mathematical framework.

The first step of the present investigation is to develop a \emph{non}-minimal
class of geometries and variables allowing manifest projective and
Lorentz invariance. This is accomplished in Section \ref{sec:Nonminimal compatibility}
where we generalize the Einstein-Cartan geometry to explicitly allow
both torsion and nonmetricity. Then a linear combination $\mathbf{S}^{a}=\mathbf{T}^{a}-\frac{1}{2}\mathbf{e}^{b}\wedge\mathbf{Q}_{ab}$
of the torsion and antisymmetric part of the nonmetricity is projectively
invariant. Remarkably, the structure equations can be expressed entirely
in terms of the torsion-like quantity $\mathbf{S}^{a}$, leading to
a Poincarè-equivalent theory with manifest projective and Lorentz
symmetries. Equally surprisingly, the new connection based on $\mathbf{S}^{a}$
is metric compatible.

From these developments, we go on to fully develop the new formulation.
Recognizing that $\mathbf{T}^{a},\mathbf{Q}^{a}\equiv\frac{1}{2}\mathbf{e}^{b}\wedge\mathbf{Q}_{\;\;\;b}^{a}$,
and $\mathbf{S}^{a}$ all lie in the vector space $\mathcal{A}_{\left[2\right]}^{1}$
of vector-valued 2-forms we easily identify the class of rotations
of torsion and nonmetricity which leave the sum $\mathbf{S}^{a}$
invariant. These rotations comprise an internal symmetry of the new
system. Because these rotations are fully decoupled from the gravitational
variables the Coleman-Mandula theorem is satisfied. While all our
calculations are carried out in arbitrary dimension $n$ and signature
$\left(p,q\right)$, we note that the 4-dim case acquires internal
symmetry $SO\left(10,9\right)$, $SO\left(11,8\right)$ or one of
the corresponding spin representations. Any of these cases is sufficient
internal symmetry for the Standard Model.

Finally, inclusion of both torsion and nonmetricity prompts reconsideration
of the gravitational action. Following a principle often used in general
relativity to justify the Einstein-Hilbert action, we write the most
general action dependent on no more than second derivatives of the
metric, and no more than linear in second derivatives. This motivates
a five--parameter addition to the Einstein-Hilbert action which includes
quadratic torsion, quadratic nonmetricity and torsion-nonmetricity
coupling terms. Imposing projective invariance reduces the additional
terms to two kinetic terms. The final action depends on $\mathbf{S}^{a}$
and curvature alone.

In Section \ref{sec:Poincar=0000E8-gauge-theory} below we lay out
basic properties of the Poincarè gauge geometry and ECSK theory, then
in Section \ref{sec:Projective-symmetry} develop projective symmetry
and its effects in Riemann-Cartan geometry. In Section \ref{sec:Nonminimal compatibility}
we carry out a revised form of the gauge construction with nonmetricity
included from the start, then show how Poincarè symmetry is recovered
by introducing the $s$-torsion $\mathbf{S}^{a}$. Section \ref{sec:Further-invariance-of}
describes the new internal rotations of the modified geometry and
in Section \ref{sec:The-action} we find the most general projectively
invariant, second order action as described above. We end with a brief
summary.

\section{Poincarè gauge theory\label{sec:Poincar=0000E8-gauge-theory}}

\subsection{General relativity and Poincarè gauge theory}

The long history of Poincarè gauging as a gauge version of general
relativity is testified by the sequence of researchers--Cartan \cite{Cartan,Cartan 1922,Cartan 1923,Cartan 1924,Cartan 1925},
Einstein \cite{Einstein 1928}, Kibble \cite{Kibble 1961}, and Sciama
\cite{Sciama1962,Sciama}--who have lent their initials to the slightly
more general ECSK theory of gravity. By adopting the Einstein-Hilbert
action, restricting the torsion to zero, and varying the metric, Poincarè
gauge theory reproduces general relativity. More generally, leaving
the torsion free and varying both the solder form and spin connection
enacts the Palatini variation in a systematic way and yields the ECSK
theory in Riemann-Cartan geometry.

Nonzero torsion introduces new features beyond general relativity.
Dirac fields couple to the totally antisymmetric part of the torsion
\cite{Datta,HehlDatta,HayashiBreegman,Hehl,HehlvonderHeydeKerlick,HehlNester,HehlNitschvonderHeyde},
while Rarita-Schwinger \cite{RaritaSchwinger,Buchdahl} and higher
spin fermions give sources to the full torsion \cite{Wheeler2023}.
While variation of the Einstein-Hilbert action limits torsion to be
nonpropagating and zero in vacuum, some authors add a dynamical term
to the action as well \cite{Neville 1980,CarrollField,SezginvanNieuwenhuizen,Saa,BelyaevShapiro}.

Torsion produces anomalous contributions to parallel transport of
any vector in a non-parallel direction. For example the evolution
of angular momentum along a timelike curve will depend on the antisymmetric
part of the connection, i.e., torsion. While the Lense-Thirring effect
of general relativity describes some effects of gravity on the propagation
of spinning objects, the change in angular momentum due to torsion
will make an additional contribution. Unfortunately, current measurements
\cite{GP-B Summary} are not precise enough to place limits on the
magnitude of torsion. The strongest limits are found in \cite{Neville 1980,Shapiro,BelyaevShapiro}.

\subsubsection{General relativity as a Lorentz gauge theory}

To see these differences clearly, recall the treatment of general
relativity as a Lorentz gauge theory, first described by Utiyama \cite{Utiyama}.
With a Lorentzian spacetime as the base manifold $\left(\mathcal{M},g\right)$,
we ask for a principal fiber bundle with Lorentz symmetry and symmetric
connection. In an orthonormal frame $\mathbf{e}^{a}$ the spin connection
$\boldsymbol{\alpha}_{\;\;\;b}^{a}$ satisfies
\begin{equation}
\mathbf{d}\mathbf{e}^{a}=\mathbf{e}^{b}\wedge\boldsymbol{\alpha}_{\;\;\;b}^{a}\label{GR solder form}
\end{equation}
and the Riemann curvature 2-form is given by
\begin{equation}
\mathbf{R}_{\;\;\;b}^{a}=\mathbf{d}\boldsymbol{\alpha}_{\;\;\;b}^{a}-\boldsymbol{\alpha}_{\;\;\;b}^{c}\wedge\boldsymbol{\alpha}_{\;\;\;c}^{a}\label{GR curvature}
\end{equation}
In a coordinate basis $\boldsymbol{\alpha}_{\;\;\;b}^{a}$ is the
Christoffel connection. Orthonormality of the basis $\left\langle \mathbf{e}^{a},\mathbf{e}^{b}\right\rangle =\eta^{ab}$
leads to the relationship $g_{\mu\nu}=e_{\mu}^{\;\;\;a}e_{\nu}^{\;\;\;b}\eta_{ab}$
between the components $\mathbf{e}^{a}=e_{\mu}^{\;\;\;a}\mathbf{d}x^{\mu}$
and the metric $g_{\mu\nu}$. The field equation for the metric is
then determined by the Einstein-Hilbert action plus the action for
any matter fields $S=\frac{1}{2}\int\mathbf{R}^{ab}\wedge\mathbf{e}^{c}\wedge\mathbf{e}^{d}e_{abcd}+\kappa S_{matter}$
and variation results in the familiar Einstein equation
\[
G_{\alpha\beta}=\kappa T_{\alpha\beta}
\]
where $G_{\alpha\beta}$ is the Einstein tensor and $T_{\alpha\beta}$
the energy tensor of the matter fields. The metric variation leads
to symmetric Einstein and energy tensors.

\subsubsection{Poincarè gauge theory of gravity}

In its broadest form, Poincarè gauge theory starts with the homogeneous
manifold $\mathcal{M}$ formed by the quotient of the Poincarè group
by its Lorentz subgroup $\mathcal{P}/\mathcal{L}=\mathcal{M}_{0}$.
The projection mapping from cosets of this quotient to $\mathcal{M}_{0}$
leads to a principal fiber bundle, effectively a copy of the Lorentz
group at each point of $\mathcal{M}_{0}$. By generalizing the manifold
and the Maurer-Cartan form of the spin connection $\boldsymbol{\omega}_{\;\;\;b}^{a}$
and solder form $\mathbf{e}^{a}$, while preserving the local Lorentz
symmetry of the principal fiber bundle, we arrive at expressions for
two 2-form fields.
\begin{eqnarray}
\boldsymbol{\mathcal{R}}_{\;\;\;b}^{a} & = & \mathbf{d}\boldsymbol{\omega}_{\;\;\;b}^{a}-\boldsymbol{\omega}_{\;\;\;b}^{c}\wedge\boldsymbol{\omega}_{\;\;\;c}^{a}\label{Curvature}\\
\mathbf{T}^{a} & = & \mathbf{d}\mathbf{e}^{a}-\mathbf{e}^{b}\wedge\boldsymbol{\omega}_{\;\;\;b}^{a}\label{Torsion}
\end{eqnarray}
These are the Riemann-Cartan curvature and the torsion. The action
may still be taken as Einstein-Hilbert plus matter, but with Riemann-Cartan
curvature scalar. If we then constrain the torsion to zero, $\boldsymbol{\omega}_{\;\;\;b}^{a}$
reduces to $\boldsymbol{\alpha}_{\;\;\;b}^{a}$, the curvature reduces
to Riemannian and the system reproduces general relativity. Without
the torsion constraint the resulting field equations still reduce
to the Einstein equation and vanishing torsion in vacuum, but many
matter sources lead to nonvanishing torsion \cite{Wheeler2023}. The
best known of these sources is the axial current of Dirac fields $\bar{\psi}\gamma^{a}\gamma_{5}\psi$
(\cite{Datta,HehlDatta,HayashiBreegman,Hehl,HehlvonderHeydeKerlick,HehlNester,HehlNitschvonderHeyde})
which leads to 
\begin{eqnarray*}
T_{\;\;\;bc}^{a} & = & \lambda\varepsilon_{\;\;\;bcd}^{a}\bar{\psi}\gamma^{d}\gamma_{5}\psi
\end{eqnarray*}
Most research on torsion has focussed on this totally antisymmetric
form. However it has been shown that the gravitino field \cite{RaritaSchwinger},
a spin-$\frac{3}{2}$ Rarita-Schwinger field present in supergravity
theories (for example, \cite{Freedman vanN Rarita Sch,Freedman1976,MacDowellMansouri,CremmerJuliaScherk,Passiaas11D}),
drives all components of torsion \cite{Wheeler2023}. Without adding
a propagaging term for torsion to the theory, torsion still vanishes
in vacuum, moving only as its source field moves.

\subsection{The structure of Riemann-Cartan geometry}

We review the formal features of Poincarè gauge theory. All results
below hold in arbitrary dimension $n=p+q$ and signature $s=p-q$
so while we continue to refer to the Poincarè group $ISO\left(3,1\right)$
and its Lorentz subgroup $SO\left(3,1\right)$ we actually work with
$\mathcal{P}=ISO\left(p,q\right)$ or $\mathcal{P}=ISpin\left(p,q\right)$
with subgroups $\mathcal{L}=SO\left(p,q\right)$ or $\mathcal{L}=Spin\left(p,q\right)$
respectively. The local Lorentz arena for general relativity in $n$
dimensions follows by setting $q=1$. 

In Appendix A we summarize the formal fiber bundle development of
Riemann-Cartan geometry. Here we give only the resulting basic properties. 

The most relevant results of this construction are the 2-form expressions
for the Riemann-Cartan curvature $\boldsymbol{\mathcal{R}}_{\;\;\;b}^{a}$
and torsion $\mathbf{T}^{a}$.
\begin{eqnarray}
\mathbf{d}\boldsymbol{\omega}_{\;\;\;b}^{a} & = & \boldsymbol{\omega}_{\;\;\;b}^{c}\wedge\boldsymbol{\omega}_{\;\;\;c}^{a}+\boldsymbol{\mathcal{R}}_{\;\;\;b}^{a}\label{Riemann-Cartan curvature}\\
\mathbf{d}\mathbf{e}^{a} & = & \mathbf{e}^{b}\wedge\boldsymbol{\omega}_{\;\;\;b}^{a}+\mathbf{T}^{a}\label{Riemann-Cartan torsion}
\end{eqnarray}
Each of these may be expanded in the orthonormal basis
\begin{eqnarray}
\boldsymbol{\mathcal{R}}_{\;\;\;b}^{a} & = & \frac{1}{2}\mathcal{R}_{\;\;\;bcd}^{a}\mathbf{e}^{c}\land\mathbf{e}^{d}\label{Horizontal curvature}\\
\mathbf{T}^{a} & = & \frac{1}{2}T_{\;\;\;bc}^{a}\mathbf{e}^{b}\land\mathbf{e}^{c}\label{Horizontal torsion}
\end{eqnarray}
In a coordinate basis $\mathbf{T}^{a}$ is given by any antisymmetric
part of the connection.

The Bianchi identities generalize to
\begin{eqnarray}
\boldsymbol{\mathcal{D}}\mathbf{T}^{a} & = & \mathbf{e}^{b}\land\boldsymbol{\mathcal{R}}_{\;\;\;b}^{a}\label{1st RC Bianchi}\\
\boldsymbol{\mathcal{D}}\boldsymbol{\mathcal{R}}_{\;\;\;b}^{a} & = & 0\label{2nd Riemann Cartan Bianchi}
\end{eqnarray}
where the covariant exterior derivatives are given by
\begin{eqnarray*}
\boldsymbol{\mathcal{D}}\boldsymbol{\mathcal{R}}_{\;\;\;b}^{a} & = & \mathbf{d}\boldsymbol{\mathcal{R}}_{\;\;\;b}^{a}+\boldsymbol{\mathcal{R}}_{\;\;\;b}^{c}\wedge\boldsymbol{\omega}_{\;\;\;c}^{a}-\boldsymbol{\omega}_{\;\;\;b}^{c}\wedge\boldsymbol{\mathcal{R}}_{\;\;\;c}^{a}\\
\boldsymbol{\mathcal{D}}\mathbf{T}^{a} & = & \mathbf{d}\mathbf{T}^{a}+\mathbf{T}^{b}\wedge\boldsymbol{\omega}_{\;\;\;b}^{a}
\end{eqnarray*}
The frame field $\mathbf{e}^{a}$ is $\left(p,q\right)$-orthonormal,
$\left\langle \mathbf{e}^{a},\mathbf{e}^{b}\right\rangle =\eta^{ab}=diag\left(1,\ldots,1,-1,\ldots,-1\right)$,
with the connection assumed to be metric compatible
\begin{eqnarray}
\mathbf{d}\eta_{ab}-\eta_{cb}\boldsymbol{\omega}_{\;\;\;a}^{c}-\eta_{ac}\boldsymbol{\omega}_{\;\;\;b}^{c} & = & 0\label{Metric compatibility}
\end{eqnarray}
Since $\mathbf{d}\eta^{ab}=0$, the spin connection is antisymmetric,
$\boldsymbol{\omega}_{ab}=-\boldsymbol{\omega}_{ba}$.

Equations (\ref{Curvature})-(\ref{2nd Riemann Cartan Bianchi}) describe
Riemann-Cartan geometry in the Cartan formalism. Note that the Riemann-Cartan
curvature, $\boldsymbol{\mathcal{R}}_{\;\;\;b}^{a}$, differs from
the Riemann curvature $\mathbf{R}_{\;\;\;b}^{a}$ by terms dependent
on the torsion. 

When the torsion vanishes, $\mathbf{T}^{a}=0$, the Riemann-Cartan
curvature $\boldsymbol{\mathcal{R}}_{\;\;\;b}^{a}$ reduces to the
Riemann curvature $\mathbf{R}_{\;\;\;b}^{a}$ and Eqs.(\ref{Curvature})
and (\ref{Torsion}) exactly reproduce the expressions for the connection
and curvature of a general Riemannian geometry. At the same time,
Eqs.(\ref{1st RC Bianchi}) and (\ref{2nd Riemann Cartan Bianchi})
reduce to the usual first and second Bianchi identities.

These results are geometric; a physical model follows when we posit
an action functional. The action may depend on the bundle tensors
$\mathbf{e}^{b},\mathbf{T}^{a},\boldsymbol{\mathcal{R}}_{\;\;\;b}^{a}$
and the invariant tensors $\eta_{ab}$ and $e_{ab\ldots d}$. To this
we may add source functionals built from any field representations
of the fiber symmetry group $\mathcal{L}$, including scalars, spinors,
vector fields, etc.

Constraining the torsion zero, specifying the Einstein-Hilbert form
of action, and varying only the solder form, the $q=1$ theory describes
general relativity as a gauge theory in $n$-dimensions. We cannot
vary the metric and connection independently because this can introduce
nonzero sources for torsion, making the $\mathbf{T}^{a}=0$ constraint
inconsistent.

Dropping the torsion constraint while retaining the Einstein-Hilbert
action gives the Einstein-Cartan-Sciama-Kibble (ECSK) theory of gravity
in Riemann-Cartan geometry. The torsion is found to depend on the
spin tensor, given by the connection variation of the source $\sigma$$_{\;\;\;ab}^{\mu}$$=\frac{\delta L}{\delta\omega_{\;\;\mu}^{ab}}$.
Without modifying the action to include dynamical torsion, the resulting
torsion survives only within matter.

The structure equations, Eqs.(\ref{Torsion}) and (\ref{Curvature}),
allow us to derive an explicit form for the connection and a reduced
form for the curvature. The result (see Appendix \ref{sec:Appendix-A:-Formal})
for the spin connection is
\begin{eqnarray}
\boldsymbol{\omega}_{\;\;\;b}^{a} & = & \boldsymbol{\alpha}_{\;\;\;b}^{a}+\mathbf{C}_{\;\;\;b}^{a}\label{Connection with torsion}
\end{eqnarray}
where $\mathbf{C}_{\;\;\;b}^{a}$ is the \emph{contorsion},
\begin{equation}
\mathbf{C}_{\;\;\;b}^{a}=\frac{1}{2}\left(T_{c\;\quad b}^{\;\;\;a}+T_{b\;\;\quad c}^{\;\;\;a}-T_{\;\;\;bc}^{a}\right)\mathbf{e}^{c}\label{Contorsion}
\end{equation}
Contorsion transforms tensorially so this form is unique. We may recover
the torsion by wedging and contracting with $\mathbf{e}^{b}$.
\begin{eqnarray*}
\mathbf{C}_{\;\;\;b}^{a}\wedge\mathbf{e}^{b} & = & \mathbf{T}^{a}
\end{eqnarray*}

The torsion now enters the curvature through the connection. Expanding
the Cartan-Riemann curvature of Eq.(\ref{Curvature}) using Eq.(\ref{Connection with torsion})
and identifying the $\boldsymbol{\alpha}$-covariant derivative, $\mathbf{D}\mathbf{C}_{\;\;\;b}^{a}=\mathbf{d}\mathbf{C}_{\;\;\;b}^{a}-\mathbf{C}_{\;\;\;b}^{c}\land\boldsymbol{\alpha}_{\;\;\;c}^{a}-\boldsymbol{\alpha}_{\;\;\;b}^{c}\land\mathbf{C}_{\;\;\;c}^{a}$
leads to
\begin{eqnarray}
\boldsymbol{\mathcal{R}}_{\;\;\;b}^{a} & = & \mathbf{R}_{\;\;\;b}^{a}+\mathbf{D}\mathbf{C}_{\;\;\;b}^{a}-\mathbf{C}_{\;\;\;b}^{c}\land\mathbf{C}_{\;\;\;c}^{a}\label{ECSK curvature}
\end{eqnarray}
This is the Riemann-Cartan curvature expressed in terms of the Riemann
curvature and the contorsion. If we contract with $\mathbf{e}^{b}$
we recover the Bianchi identity. This happens because our solution
for the connection automatically satisfies the integrability condition
for the connection.

\section{Projective symmetry \label{sec:Projective-symmetry}}

In this Section we review the derivation of projective transformation
of the connection in affine (nonmetric) geometry by reparameterization
of autoparallels. Then we carry out the derivation of the same transformation
in a geometry with both metric and connection, resulting from reparameterization
of geodesics. In the third Subsection we show the relationship between
projective transformation and the Weyl vector, and how extending to
a Weyl geometry creates manifest invariance of induced reparameterizations.
In the final Subsection we examine the effect of projective transformation
on the torsion in $\mathcal{P}/\mathcal{L}$ gauge theory. 

\subsection{Projective symmetry in nonmetric geometry}

Consider a principal fiber bundle with Lorentz fibers and base manifold
$\mathcal{M}$. Given a local Lorentz connection, but no metric, we
are able to define the curvature of $\mathcal{M}$ by the single Cartan
equation
\begin{eqnarray*}
\mathbf{d}\boldsymbol{\omega}_{\;\;\;b}^{a} & = & \boldsymbol{\omega}_{\;\;\;b}^{c}\wedge\boldsymbol{\omega}_{\;\;\;c}^{a}+\mathbf{R}_{\;\;\;b}^{a}
\end{eqnarray*}
Here Latin indices refer to a general basis $\mathbf{e}^{a}=e_{\alpha}^{\;\;\;a}\mathbf{d}x^{\alpha}$.
The coefficients $e_{\alpha}^{\;\;\;a}$ must be invertible, but we
cannot claim the basis forms $\mathbf{e}^{a}$ to be orthonormal.

While there are no geodesics without a metric, we may consider autoparallels.
\begin{equation}
v^{b}D_{b}v^{a}=v^{b}\partial_{b}v^{a}+\omega_{\;\;\;bc}^{a}v^{b}v^{c}=0\label{Autoparallel}
\end{equation}
Projective transformations are changes of the connection that leave
the curvature and autoparallels invariant. They arise from reprameterization
of the autoparallels. 

Let $v^{a}=e_{\alpha}^{\;\;\;a}\frac{dx^{\alpha}}{d\lambda}$ be tangent
to the autoparallel curve, and consider a reparameterization $\sigma=\sigma\left(\lambda\right)$
to a parallel vector $u^{\alpha}$
\[
v^{a}=\frac{d\sigma}{d\lambda}\frac{dx^{a}}{d\sigma}=\frac{d\sigma}{d\lambda}u^{a}
\]
where $\sigma\left(\lambda\right)$ is monotonic. In order for the
curvature and second Bianchi identity to remain continuous $\sigma\left(\lambda\right)$
should have at least up to third derivatives. Let $f=\frac{d\sigma}{d\lambda}$
and substitute for $v^{a}$ in Eq.(\ref{Autoparallel}) to find
\begin{equation}
u^{b}\partial_{b}u^{a}+\left(\omega_{\;\;\;bc}^{a}+\delta_{c}^{a}\partial_{b}\left(\ln f\right)\right)u^{b}u^{c}=0\label{Reparameterized autoparallels}
\end{equation}
From this we extract the transformed connection
\begin{eqnarray*}
\tilde{\omega}_{\;\;\;bc}^{a} & = & \omega_{\;\;\;bc}^{a}+\delta_{c}^{a}\partial_{b}\left(\ln f\right)
\end{eqnarray*}
Notice that the projective change in the connection could be symmetrized,
$\omega_{\;\;\;bc}^{a}+\delta_{(c}^{a}\partial_{b)}\left(\ln f\right)$,
when we remove $u^{b}u^{c}$ but this does not preserve the curvature.

Setting $\boldsymbol{\xi}=\mathbf{d}\ln f$, the we recover the original
form of (\ref{Autoparallel}) in terms of $\tilde{\boldsymbol{\omega}}_{\;\;\;b}^{a}$
if we write
\begin{eqnarray}
\tilde{\boldsymbol{\omega}}_{\;\;\;b}^{a} & = & \boldsymbol{\omega}_{\;\;\;b}^{a}+\delta_{b}^{a}\boldsymbol{\xi}\nonumber \\
\mathbf{d}\boldsymbol{\xi} & = & 0\label{Projective transformation}
\end{eqnarray}
 This is the projective transformation of the connection.

The invariance of the curvature is immediate.
\begin{eqnarray*}
\mathbf{\tilde{R}}_{\;\;\;b}^{a} & = & \mathbf{d}\tilde{\boldsymbol{\omega}}_{\;\;\;b}^{a}-\tilde{\boldsymbol{\omega}}_{\;\;\;b}^{c}\wedge\tilde{\boldsymbol{\omega}}_{\;\;\;c}^{a}\\
 & = & \mathbf{d}\left(\boldsymbol{\omega}_{\;\;\;b}^{a}+\delta_{b}^{a}\boldsymbol{\xi}\right)-\left(\boldsymbol{\omega}_{\;\;\;b}^{c}+\delta_{b}^{c}\boldsymbol{\xi}\right)\wedge\left(\boldsymbol{\omega}_{\;\;\;c}^{a}+\delta_{c}^{a}\boldsymbol{\xi}\right)\\
 & = & \mathbf{d}\boldsymbol{\omega}_{\;\;\;b}^{a}-\boldsymbol{\omega}_{\;\;\;b}^{c}\wedge\boldsymbol{\omega}_{\;\;\;c}^{a}+\delta_{b}^{a}\mathbf{d}\boldsymbol{\xi}
\end{eqnarray*}
Since $\mathbf{d}\boldsymbol{\xi}=0$, the curvature is unchanged,
$\tilde{\mathbf{R}}_{\;\;\;b}^{a}=\mathbf{R}_{\;\;\;b}^{a}$ and no
new structures are introduced.

If we had symmetrized when stripping the tangent vectors off of Eq.(\ref{Reparameterized autoparallels})
we instead find $\tilde{\boldsymbol{\omega}}_{\;\;\;b}^{a}=\boldsymbol{\omega}_{\;\;\;b}^{a}+\frac{1}{2}\left(\delta_{b}^{a}\xi_{\alpha}+\delta_{\alpha}^{a}\xi_{b}\right)\mathbf{d}x^{\alpha}$
where $\mathbf{d}x^{\alpha}$ is a coordinate basis on $\mathcal{M}$.
With vanishing torsion and $\mathbf{d}\boldsymbol{\xi}=0$ the curvature
now changes to
\begin{eqnarray*}
\mathbf{\tilde{R}}_{\;\;\;b}^{a} & = & \mathbf{R}_{\;\;\;b}^{a}+\frac{1}{2}\delta_{\alpha}^{a}\left(\mathbf{D}\xi_{b}-\xi_{b}\boldsymbol{\xi}\right)\wedge\mathbf{d}x^{\alpha}
\end{eqnarray*}

In the next subsection we show that when we have a metric the corresponding
projective transformation also leaves geodesics invariant.

\subsection{Projective symmetry of geodesics}

The situation within Poincarè gauge theory is different from the affine
case. Here the orthonormal frame fields provide a metric, so we may
consider geodesics instead of autoparallels. At the same time, projective
symmetry produces additional, non-invariant changes.

The structure equations are now those of Eqs.(\ref{Curvature}) and
(\ref{Torsion}) with the spin connection appropriate to signature
$\left(p,q\right)$ symmetry.

Let the proper length of a curve $x^{\alpha}\left(\lambda\right)$
be given by $s=\intop\sqrt{\kappa\eta_{ab}v^{a}v^{b}}d\lambda$ where
$\lambda$ is an arbitrary parameterization for tangent vectors $v^{a}=e_{\alpha}^{\;\;\;a}\frac{dx^{\alpha}}{d\lambda}$.
The curve is spacelike or timelike for $\kappa=\pm1$, respectively.
Varying the arclength $s\left[x\right]$ with respect to the curve
$x^{\alpha}\left(\lambda\right)$ with arbitrary parameterization
$\lambda$ leads to
\begin{eqnarray*}
\frac{dv^{\nu}}{d\lambda} & = & -\Gamma_{\;\;\;\alpha\beta}^{\nu}v^{\alpha}v^{\beta}+\frac{1}{2}\frac{1}{\left|v^{2}\right|}\left(\frac{d}{d\lambda}\left|v^{2}\right|\right)v^{\nu}
\end{eqnarray*}
where $\Gamma_{\;\;\;\alpha\beta}^{\nu}$ is the Christoffel connection
and $\left|v^{2}\right|=\kappa\eta_{ab}v^{a}v^{b}$.

Since we now have a preferred parameterization by proper time
\[
u^{\alpha}=\frac{dx^{\alpha}}{d\tau}
\]
we may refer alternate parameterizations to $\tau$.
\begin{eqnarray*}
v^{\alpha}=\frac{dx^{\alpha}}{d\lambda} & = & \frac{1}{f}u^{\alpha}
\end{eqnarray*}
where $f=\frac{1}{c}\frac{d\lambda}{d\tau}$. It follows that $\left|v^{2}\right|=-\frac{\kappa}{f^{2}}$
and the geodesic equation becomes
\[
\frac{dv^{\nu}}{d\lambda}=-\left(\Gamma_{\;\;\;\alpha\beta}^{\nu}+\delta_{\alpha}^{\nu}\partial_{\beta}\ln f\right)v^{\alpha}v^{\beta}
\]
Returning to the spin connection, the projective transformation is

\begin{eqnarray*}
\tilde{\boldsymbol{\alpha}}_{\;\;\;b}^{a} & = & \boldsymbol{\alpha}_{\;\;\;b}^{a}+\delta_{b}^{a}\boldsymbol{\xi}\\
\boldsymbol{\xi} & = & \mathbf{d}\left(\ln f\right)
\end{eqnarray*}
in agreement with Eq.(\ref{Projective transformation}).

It is important to demonstrate that $\xi_{\nu}$ is a well-defined
field. We know that $v^{2}$ is a function of $\lambda$ for any curve,
but we need to verify that $\lambda\left(x^{\alpha}\right)$ is a
differentiable function. The sketch of a proof follows.

Suppose we start at a fixed point $\mathcal{P}$ and consider curves
through $\mathcal{P}$, with parameterizations such that $x^{\alpha}\left(\lambda=0\right)=\mathcal{P}$.
Let $\mathcal{Q}$ be a second point and consider curves passing through
both $\mathcal{P}$ and $\mathcal{Q}$. Then there is no single value
of $\lambda$ at $\mathcal{Q}$, since the curves will have differing
proper length. However, for nearby $\mathcal{P},\mathcal{Q}$ there
is a unique geodesic $x_{0}^{\alpha}\left(\lambda\right)$ and we
may assign the value $\lambda\left(\mathcal{Q}\right)$ as the parameter
value which the geodesic parameter attains at $\mathcal{Q}$, i.e.,
$x_{0}^{\alpha}\left(\lambda\left(\mathcal{Q}\right)\right)=\mathcal{Q}$.
Now suppose $\mathcal{P},\mathcal{Q}$ are points connected by multiple
geodesics (e.g., the north and south poles of a sphere). Then these
geodesics must yield the same value of $\lambda$, or else there is
a minimum value of $\lambda$ (e.g., curves around a cylinder in opposite
directions, with $\mathcal{P},\mathcal{Q}$ nearer in one of the directions).
We take this minimum for the value of $\lambda\left(\mathcal{Q}\right)$.
This gives a unique value $\lambda\left(x^{\alpha}\right)$ to each
point $\mathcal{Q}$ that can be reached from $\mathcal{P}$. The
extremal condition requires small changes in the path to produce small
changes in the proper length of any curve, and these only at second
order. Therefore, the function is differentiable.

We now have an equivalence class of connections,
\begin{equation}
\tilde{\Gamma}_{\;\;\;\mu\nu}^{\alpha}\in\left\{ \Gamma_{\;\;\;\mu\nu}^{\alpha}-\delta_{\mu}^{\alpha}\xi_{\nu}|\mathbf{d}\boldsymbol{\xi}=0\right\} \label{Equivalence class}
\end{equation}
As we have seen, projective transformations preserve the curvature
and therefore the action and field equations.

\subsection{Minimal compatibility}

We define \emph{minimal compatibility} to mean the minimum change
in the geometry required to achieve manifest reparameterization and
local Lorentz invariance. This compatibility is implicit in the Ehlers,
Pirani, Schild program \cite{EPS,Matveev1,Matveev2}. Here we show
how extending to an integrable Weyl geometry achieves reparameteriation
invariance.

General relativity has both metric and connection, satisfying Eqs.(\ref{GR curvature})
and (\ref{GR solder form}). but Eq.(\ref{GR solder form})) changes
with projective transformation to give
\begin{eqnarray}
\mathbf{d}\mathbf{e}^{a} & = & \mathbf{e}^{b}\wedge\boldsymbol{\alpha}_{\;\;\;b}^{a}+\mathbf{e}^{a}\wedge\boldsymbol{\xi}\label{Transformed solder equation}
\end{eqnarray}
The connection is no longer fully compatible with the metric, but
gives it a nonvanishing covariant derivative
\begin{eqnarray}
\mathbf{Q}_{ab} & = & \mathbf{d}\eta_{ab}-\eta_{cb}\left(\boldsymbol{\alpha}_{\;\;\;a}^{c}+\delta_{a}^{c}\boldsymbol{\xi}\right)-\eta_{ac}\left(\boldsymbol{\alpha}_{\;\;\;b}^{c}+\delta_{b}^{c}\boldsymbol{\xi}\right)\nonumber \\
 & = & -2\eta_{ab}\boldsymbol{\xi}\label{Projective change in nonmetricity}
\end{eqnarray}
Here the trace of $\mathbf{Q}_{ab}$ is proportional to $\boldsymbol{\xi}$.

We show that this trace is proportional the the Weyl vector. Notice
that the non-metricity $Q_{abc}=D_{c}g_{ab}$ changes if a different
metric is chosen, so that the metric becomes a gauge choice for $Q_{abc}$.
In particular, if we change $g_{ab}$ by a conformal factor, $\tilde{g}_{ab}=e^{2\varphi}g_{ab}$
the non-metricity changes to
\[
\tilde{\mathbf{Q}}_{ab}=\mathbf{D}\tilde{g}_{ab}=e^{2\phi}\mathbf{Q}_{ab}+2\tilde{g}_{ab}\mathbf{d}\phi
\]
Suppose the non-metricity is pure trace, $Q_{\;\;\;bc}^{a}=\frac{1}{n}\delta_{b}^{a}\sigma_{c}$.
Then the non-metric connection becomes
\begin{eqnarray*}
\omega_{abc} & = & \alpha_{abc}-\frac{1}{2}\left(Q_{abc}+Q_{cab}-Q_{bca}\right)\\
 & = & \alpha_{abc}-\frac{1}{2}\left(\frac{1}{n}\eta_{ab}\sigma_{c}+\frac{1}{n}\eta_{ca}\sigma_{b}-\frac{1}{n}\eta_{bc}\sigma_{a}\right)
\end{eqnarray*}
With the identifications $W_{a}=-\frac{1}{2n}\sigma_{a}$ and $Q_{\;\;\;ac}^{a}=-2nW_{c}$
this has the form of the connection of a Weyl geometry. Moreover,
since a conformal change changes the non-metricity as
\begin{eqnarray*}
\tilde{Q}_{abc} & = & e^{2\phi}Q_{abc}+\tilde{g}_{ab}\phi_{,c}
\end{eqnarray*}
the trace changes by the correct gauge transformation
\begin{eqnarray*}
\tilde{W}_{c} & = & W_{c}-\phi_{,c}
\end{eqnarray*}
and we may identify the trace of non-metricity with the Weyl vector.

Making the substitution $\boldsymbol{\omega}=W_{a}\mathbf{e}^{a}=-\boldsymbol{\xi}$
puts the structure equations in the form
\begin{eqnarray*}
\mathbf{d}\boldsymbol{\omega}_{\;\;\;b}^{a} & = & \boldsymbol{\omega}_{\;\;\;b}^{c}\wedge\boldsymbol{\omega}_{\;\;\;c}^{a}+\boldsymbol{\mathcal{R}}_{\;\;b}^{a}\\
\mathbf{d}\mathbf{e}^{a} & = & \mathbf{e}^{b}\wedge\boldsymbol{\omega}_{\;\;\;b}^{a}+\boldsymbol{\omega}\wedge\mathbf{e}^{a}
\end{eqnarray*}
The connection is now that of a Weyl geometry \cite{Wheeler2018}.
This reflects multiple changes. The antisymmetry of the connection
is restored and the Weyl connection is metric compatible. However,
the connection and curvature now include contributions from the Weyl
vector.
\begin{eqnarray*}
\boldsymbol{\omega}_{\;\;\;b}^{a} & = & \boldsymbol{\alpha}_{\;\;\;b}^{a}+W_{b}\mathbf{e}^{a}-W^{a}\eta_{bc}\mathbf{e}^{c}\\
\boldsymbol{\mathcal{R}}_{\;\;b}^{a} & = & \mathbf{R}_{\;\;b}^{a}-\left(\delta_{d}^{a}\delta_{b}^{c}-\eta^{ac}\eta_{bd}\right)\left(\mathbf{D}W_{c}+W_{c}\boldsymbol{\omega}\right)\wedge\mathbf{e}^{d}
\end{eqnarray*}
These extra contributions make both fields invariant under dilatations,
\begin{eqnarray*}
\tilde{\boldsymbol{\omega}} & = & \boldsymbol{\omega}+\mathbf{d}\phi\\
\mathbf{e}^{a} & = & e^{\phi}\mathbf{e}^{a}
\end{eqnarray*}
Moreover, since the Weyl geometry is integrable there is a choice
of $e^{\phi}$ with vanishing Weyl vector, restoring the Riemannian
form $\mathbf{R}_{\;\;b}^{a}$.

Dilatations induce reparameterizations of curves. Concretely, the
line element rescales as
\[
d\tilde{s}^{2}=e^{2\phi}ds^{2}
\]
so that along a curve $x^{\alpha}\left(\tau\right)$ a tangent vector
rescales to
\[
\tilde{u}^{a}=e_{\alpha}^{\;\;\;a}\frac{dx^{\alpha}}{d\tau}\rightarrow e^{\phi}u^{a}
\]
If we set $f=\frac{d\lambda}{d\tau}=e^{\phi}$ this is equivalent
to a reparameterization
\[
\lambda\left(\tau\right)=\intop_{0}^{\tau}e^{\phi}d\tau
\]
By realizing $f\left(\tau\right)=\frac{d\lambda}{d\tau}$ as a function
$f\left(x\right)$, the reparameterization of all geodesics is equivalent
to a rescaling of the spacetime metric by $e^{2\phi}$, where $\boldsymbol{\xi}=\mathbf{d}\ln f=\mathbf{d}\ln e^{\phi}=\mathbf{d}\phi$.

\subsection{Projective symmetry in (p,q) gauge theory}

The principal difference between general relativity and Riemann-Cartan
geometry is the presence of torsion. Substituting a projective transformation
(\ref{Projective transformation}) into the equations for the curvature
and torsion, the curvature is unchanged but the torsion is altered.
\begin{eqnarray*}
\tilde{\boldsymbol{\mathcal{R}}}_{\;\;\;b}^{a} & = & \boldsymbol{\mathcal{R}}_{\;\;\;b}^{a}\\
\tilde{\mathbf{T}}^{a} & = & \mathbf{d}\mathbf{e}^{a}-\mathbf{e}^{b}\wedge\boldsymbol{\omega}_{\;\;\;b}^{a}-\mathbf{e}^{a}\wedge\boldsymbol{\xi}\\
 & = & \mathbf{T}^{a}-\mathbf{e}^{a}\wedge\boldsymbol{\xi}
\end{eqnarray*}
Just as we found for general relativity, the nonmetricity changes
as well
\begin{eqnarray}
\tilde{\mathbf{Q}}_{ab} & = & \mathbf{Q}_{ab}-2\eta_{ab}\boldsymbol{\xi}\label{Nonmetricity change}
\end{eqnarray}
Minimal compatibility, replacing $\boldsymbol{\xi}=-\boldsymbol{\omega}$
to form a Weyl geometry, again leads to simultaneously projective
and local Lorentz invariant forms.

However, there is a nonminimal approach that becomes evident when
we include both nonmetricity and torsion from the start. Before defining
nonminimal compatibility, we explicitly include nonmetricity in $\left(p,q\right)$
gauge theory.

\section{Nonminimal compatibility \label{sec:Nonminimal compatibility}}

Projective transformations alter both the torsion and the nonmetricity
of $\left(p,q\right)$ gauge theory, and we showed how minimal compatibility
restores invariance of the torsion and curvature while eliminating
the nonmetricity. Next, we revise the description of the geometry
given in Section (\ref{sec:Poincar=0000E8-gauge-theory}), explicitly,
but in a restricted way, by including nonmetricity\footnote{We stress that our approach is \emph{not} that of metric-affine gravity
\cite{HehlMetricAffine}, $f\left(R\right)$ gravity or any of several
other alternative gravity theories based on a $GL\left(n\right)$
connection. We retain Lorentz structure.}. Since nonmetricity is a tensor under local Lorentz transformations
we may introduce it while ultimately requiring no modification of
the $\mathcal{P}/\mathcal{L}$ fiber bundle.

\subsection{Revisiting the Poincarè structure equations}

Once again carrying out the Cartan procedure described in Appendix
A we develop a principal fiber bundle with $SO\left(p,q\right)$ symmetry.
This time, we drop the assumption of metric compatibility and elevate
the metric compatibility condition to the level of the structure equations.
\begin{eqnarray*}
\mathbf{d}\tilde{\boldsymbol{\omega}}_{\;\;\;b}^{a} & = & \tilde{\boldsymbol{\omega}}_{\;\;\;b}^{c}\wedge\tilde{\boldsymbol{\omega}}_{\;\;\;c}^{a}\\
\mathbf{d}\tilde{\mathbf{e}}^{a} & = & \tilde{\mathbf{e}}^{b}\wedge\tilde{\boldsymbol{\omega}}_{\;\;\;b}^{a}\\
\mathbf{d}\eta_{ab} & = & \eta_{cb}\tilde{\boldsymbol{\omega}}_{\;\;\;a}^{c}+\eta_{ac}\tilde{\boldsymbol{\omega}}_{\;\;\;b}^{c}
\end{eqnarray*}
When we modify the solder form and the spin connection $\left(\tilde{\mathbf{e}}^{b},\tilde{\boldsymbol{\omega}}_{\;\;\;b}^{a}\right)\rightarrow\left(\mathbf{e}^{b},\boldsymbol{\omega}_{\;\;\;b}^{a}\right)$,
Eqs. (\ref{Curvature}) and (\ref{Torsion}) are augmented by a third
tensor field, the 1-form nonmetricity $\mathbf{Q}^{ab}$. The presence
of $\mathbf{Q}^{ab}$ modifies the curvature $\boldsymbol{\mathcal{R}}_{\;\;\;b}^{a}$
and torsion $\mathbf{T}^{a}$, through its effect on the spin connection. 

\begin{eqnarray}
\mathbf{d}\boldsymbol{\omega}_{\;\;\;b}^{a} & = & \boldsymbol{\omega}_{\;\;\;b}^{c}\wedge\boldsymbol{\omega}_{\;\;\;c}^{a}+\boldsymbol{\mathcal{R}}_{\;\;\;b}^{a}\label{New Curvature}\\
\mathbf{d}\mathbf{e}^{a} & = & \mathbf{e}^{b}\wedge\boldsymbol{\omega}_{\;\;\;b}^{a}+\mathbf{T}^{a}\label{New Torsion}\\
\mathbf{d}\eta_{ab} & = & \eta_{cb}\boldsymbol{\omega}_{\;\;\;a}^{c}+\eta_{ac}\boldsymbol{\omega}_{\;\;\;b}^{c}+\mathbf{Q}_{ab}\label{New Nonmetricity}
\end{eqnarray}
Each new Lorentz tensor is horizontal
\begin{eqnarray*}
\boldsymbol{\mathcal{R}}_{\;\;\;b}^{a} & = & \frac{1}{2}\mathcal{R}_{\;\;\;bcd}^{a}\mathbf{e}^{c}\land\mathbf{e}^{d}\\
\mathbf{T}^{a} & = & \frac{1}{2}T_{\;\;\;bc}^{a}\mathbf{e}^{b}\land\mathbf{e}^{c}\\
\mathbf{Q}_{ab} & = & Q_{abc}\mathbf{e}^{c}
\end{eqnarray*}
to preserve the local $SO\left(p,q\right)$ symmetry.

The Bianchi identities now take the form
\begin{eqnarray*}
\boldsymbol{\mathcal{D}}\boldsymbol{\mathcal{R}}_{\;\;\;b}^{a} & = & 0\\
\boldsymbol{\mathcal{D}}\mathbf{T}^{a} & = & \mathbf{e}^{b}\land\boldsymbol{\mathcal{R}}_{\;\;\;b}^{a}\\
\boldsymbol{\mathcal{D}}\mathbf{Q}_{ab} & = & -\boldsymbol{\mathcal{R}}_{ab}-\boldsymbol{\mathcal{R}}_{ba}
\end{eqnarray*}
with the covariant derivatives defined by
\begin{eqnarray*}
\boldsymbol{\mathcal{D}}\boldsymbol{\mathcal{R}}_{\;\;\;b}^{a} & = & \mathbf{d}\boldsymbol{\mathcal{R}}_{\;\;\;b}^{a}+\boldsymbol{\mathcal{R}}_{\;\;\;b}^{c}\wedge\boldsymbol{\omega}_{\;\;\;c}^{a}-\boldsymbol{\omega}_{\;\;\;b}^{c}\wedge\boldsymbol{\mathcal{R}}_{\;\;\;c}^{a}\\
\boldsymbol{\mathcal{D}}\mathbf{T}^{a} & = & \mathbf{d}\mathbf{T}^{a}+\mathbf{T}^{b}\wedge\boldsymbol{\omega}_{\;\;\;b}^{a}\\
\boldsymbol{\mathcal{D}}\mathbf{Q}_{ab} & = & \mathbf{d}\mathbf{Q}_{ab}+\mathbf{Q}_{cb}\wedge\boldsymbol{\omega}_{\;\;\;a}^{c}+\mathbf{Q}_{ac}\wedge\boldsymbol{\omega}_{\;\;\;b}^{c}
\end{eqnarray*}
The plus signs in the derivative of nonmetricity occur because $\mathbf{Q}_{ab}$
is a 1-form.

Equations (\ref{New Curvature})-(\ref{New Nonmetricity}) describe
$SO\left(p,q\right)$ covariant tensors $\boldsymbol{\mathcal{R}}_{\;\;\;b}^{a},\mathbf{T}^{a},\mathbf{Q}_{ab}$.
Unlike a full $GL\left(n\right)$ connection, the inhomogeneous part
of local $SO\left(p,q\right)$ transformations of the connection is
antisymmetric, hence an element of the Lorentz Lie algebra.

\subsection{Solving for the connection}

This form of the structure equations is sufficient to lead to the
well-known explicit expressions for the connection and curvature.
Starting from the Riemann-Cartan connection Eq.(\ref{Connection with torsion})
we add a third term
\begin{equation}
\boldsymbol{\omega}_{\;\;\;b}^{a}=\boldsymbol{\alpha}_{\;\;\;b}^{a}+\mathbf{C}_{\;\;\;b}^{a}+\mathbf{E}_{\;\;\;b}^{a}\label{General connection}
\end{equation}
which must satisfy both Eq.(\ref{New Torsion}) and from constancy
$\mathbf{d}\eta_{ab}=0$ of the $\left(p,q\right)$ metric
\[
\mathbf{Q}_{ab}=-\boldsymbol{\omega}_{ab}-\boldsymbol{\omega}_{ba}=-\mathbf{E}_{ab}-\mathbf{E}_{ba}
\]
The torsion equation (\ref{New Torsion}) implies $\mathbf{e}^{b}\wedge\mathbf{E}_{\;\;\;b}^{a}=0$
so the pair of conditions together require
\begin{eqnarray*}
E_{abc}+E_{bac} & = & -Q_{abc}\\
E_{abc}-E_{acb} & = & 0
\end{eqnarray*}
Cycling indices of the first and combining in the usual way $\left(++-\right)$
using the second we find
\begin{eqnarray}
E_{abc} & = & -\frac{1}{2}\left(Q_{abc}+Q_{cab}-Q_{bca}\right)\label{Contricity}
\end{eqnarray}
with the connection given by Eq.(\ref{General connection}). We note
that $E_{abc}=E_{acb}$, and this insures that $\mathbf{e}^{b}\land\mathbf{E}_{\;\;\;b}^{a}=0$.
There is no change in the fiber bundle structure.

\subsection{Nonminimal compatibility}

\subsubsection{Hints at a symmetry}

A certain symmetry between torsion and nonmetricity is occasionally
noted. This generally stems from an ambiguity in the solder form structure
equation in Weyl geometry with torsion.

\begin{equation}
\mathbf{d}\mathbf{e}^{a}=\mathbf{e}^{b}\wedge\boldsymbol{\omega}_{\;\;\;b}^{a}+\boldsymbol{\omega}\wedge\mathbf{e}^{a}+\mathbf{T}^{a}\label{Weyl geom with torsion}
\end{equation}
While the Weyl vector was first introduced as a form of nonmetricity
\begin{eqnarray*}
\mathbf{D}g_{\alpha\beta} & = & \boldsymbol{\omega}g_{\alpha\beta}
\end{eqnarray*}
the extra term $\boldsymbol{\omega}\wedge\mathbf{e}^{a}$ in Eq.(\ref{Weyl geom with torsion})
can also be absorbed into the torsion
\[
\tilde{\mathbf{T}}^{a}=\mathbf{T}^{a}+\boldsymbol{\omega}\wedge\mathbf{e}^{a}
\]
Of course, since dilatational gauging induces projective transformations,
this duality is seen for projective transformations as well (for a
recent example, see \cite{IosifidisQTsymmetry}).

This dual role is also evident in the basis dependence of torsion
and nonmetricity.

The torsion is algebraic in a coordinate basis. The inhomogeneous
change in the Christoffel connection under diffeomorphisms is symmetric,
so any antisymmetric part of the connection is a tensor
\[
T_{\;\;\;\mu\nu}^{\alpha}=\Gamma_{\;\;\;\nu\mu}^{\alpha}-\Gamma_{\;\;\;\mu\nu}^{\alpha}
\]
The nonmetricity in the same basis is differential, given by the covariant
derivative of the metric
\[
Q_{\alpha\beta\mu}=D_{\mu}g_{\alpha\beta}
\]

This situation is reversed in an orthonormal basis, $\mathbf{e}^{a}$.
The torsion becomes the covariant exterior derivative of the solder
form
\begin{eqnarray*}
\mathbf{T}^{a} & = & \mathbf{D}\mathbf{e}^{a}
\end{eqnarray*}
while the nonmetricity is algebraic
\[
\mathbf{Q}_{ab}=\mathbf{D}\eta_{ab}=-\boldsymbol{\omega}_{ab}-\boldsymbol{\omega}_{ba}
\]
Since an infinitesimal local Lorentz transformation is antisymmetric,
the inhomogeneous change in the spin connection is antisymmetric and
the symmetric part is a tensor. Torsion and nonmetricity have exchanged
roles.

\subsubsection{A new projective invariant}

We find that the simplicity of these somewhat vague observations stems
from a much broader overlap between nonmetricity and torsion. Under
projective transformation (\ref{Projective transformation}) the tensors
of Eqs.(\ref{New Curvature})-(\ref{New Nonmetricity}) become
\begin{eqnarray}
\boldsymbol{\tilde{\mathcal{R}}}_{\;\;\;b}^{a} & = & \boldsymbol{\mathcal{R}}_{\;\;\;b}^{a}\label{Projective invariance of curvature}\\
\tilde{\mathbf{T}}^{a} & = & \mathbf{T}^{a}-\mathbf{e}^{a}\land\boldsymbol{\xi}\label{Projective transformation of torsion}\\
\tilde{\mathbf{Q}}_{ab} & = & \mathbf{Q}_{ab}-2\eta_{ab}\boldsymbol{\xi}\label{Projective transformation of nonmetricity}
\end{eqnarray}

The changes produced by projective transformations in both torsion
(\ref{Projective transformation of torsion}) and nonmetricity (\ref{Projective transformation of nonmetricity})
allow us to construct a projectively invariant tensor. The irreducible
parts of nonmetricity are totally symmetric $Q_{\left(abc\right)}$
and mixed symmetry $Q_{a\left[bc\right]}$. The vector space spanned
by $Q_{a\left[bc\right]}$ includes the non-totally-symmetric pieces
$Q_{a\left(bc\right)}-Q_{\left(abc\right)}$. We separate the mixed
symmetry part by writing the 2-form 
\begin{eqnarray*}
\mathbf{Q}^{a} & \equiv & \frac{1}{2}\mathbf{e}^{c}\land\mathbf{Q}_{\;\;\;c}^{a}\\
 & = & \frac{1}{2}\eta^{ab}\mathbf{e}^{c}\land\mathbf{Q}_{bc}
\end{eqnarray*}
From Eq.(\ref{Projective transformation of nonmetricity}) transforms
as
\begin{eqnarray*}
\tilde{\mathbf{Q}}^{a} & = & \mathbf{Q}^{a}-\mathbf{e}^{a}\land\boldsymbol{\xi}
\end{eqnarray*}
It follows that the combination
\begin{eqnarray}
\mathbf{S}^{a} & \equiv & \mathbf{T}^{a}-\mathbf{Q}^{a}\label{QT coupling}
\end{eqnarray}
is projectively invariant.

\subsection{The spin connection in terms of $\mathbf{S}^{a}$}

The component expansion
\begin{eqnarray*}
S_{abc} & = & T_{abc}-\frac{1}{2}\left(Q_{abc}-Q_{acb}\right)
\end{eqnarray*}
allows us to solve for the torsion in terms of $S_{abc}$ and the
mixed symmetry part of $Q_{abc}$. Substituting to eliminate the torsion

\begin{eqnarray*}
\mathbf{C}_{ab}+\mathbf{E}_{ab} & = & \frac{1}{2}\left(T_{cab}+T_{bac}-T_{abc}\right)\mathbf{e}^{c}-\frac{1}{2}\left(Q_{abc}+Q_{cab}-Q_{bca}\right)\mathbf{e}^{c}\\
 & = & \frac{1}{2}\left(S_{cab}+\frac{1}{2}\left(Q_{cab}-Q_{cba}\right)+S_{bac}+\frac{1}{2}\left(Q_{bac}-Q_{bca}\right)-S_{abc}-\frac{1}{2}\left(Q_{abc}-Q_{acb}\right)\right)\mathbf{e}^{c}\\
 &  & -\frac{1}{2}\left(Q_{abc}+Q_{cab}-Q_{bca}\right)\mathbf{e}^{c}\\
 & = & \frac{1}{2}\left(S_{cab}+S_{bac}-S_{abc}\right)\mathbf{e}^{c}-\frac{1}{2}Q_{abc}\mathbf{e}^{c}
\end{eqnarray*}
where somewhat surprisingly all but one of the nonmetricity terms
cancel. Define
\begin{eqnarray}
\mathbf{C}_{ab}^{\left(S\right)} & \equiv & \frac{1}{2}\left(S_{cab}+S_{bac}-S_{abc}\right)\mathbf{e}^{c}\label{Constortsion}
\end{eqnarray}
This is the contorsion tensor of $S_{cab}$. Finally, with
\begin{eqnarray}
\boldsymbol{\omega}_{\;\;\;b}^{a} & = & \boldsymbol{\alpha}_{\;\;\;b}^{a}+\mathbf{C}_{\;\;\;b}^{a}+\mathbf{E}_{\;\;\;b}^{a}\nonumber \\
 & = & \boldsymbol{\alpha}_{\;\;\;b}^{a}+\mathbf{C}_{\;\;\;\quad b}^{\left(S\right)a}-\frac{1}{2}Q_{\;\;\;bc}^{a}\mathbf{e}^{c}\label{Original connection}
\end{eqnarray}
the structure equation Eq.(\ref{New Torsion}) becomes
\begin{eqnarray*}
\mathbf{d}\mathbf{e}^{a} & = & \mathbf{e}^{b}\wedge\boldsymbol{\alpha}_{\;\;\;b}^{a}+\mathbf{e}^{b}\wedge\mathbf{C}_{\;\;\quad b}^{\left(S\right)a}+\mathbf{T}^{a}-\frac{1}{2}\mathbf{e}^{b}\wedge\mathbf{Q}_{ab}\\
 & = & \mathbf{e}^{b}\wedge\boldsymbol{\alpha}_{\;\;\;b}^{a}+\mathbf{e}^{b}\wedge\mathbf{C}_{\;\;\quad b}^{\left(S\right)a}+\mathbf{S}^{a}
\end{eqnarray*}
The connection from this form of the structure equation will now be
\begin{eqnarray}
\boldsymbol{\omega}_{\left(S\right)\;b}^{a} & = & \boldsymbol{\alpha}_{\;\;\;b}^{a}+\mathbf{C}_{\;\;\quad b}^{\left(S\right)a}\label{Projectively invariant connection}
\end{eqnarray}
and we recover the form of the original $\left(p,q\right)$ structure
equations with the projectively invariant $s$-torsion
\begin{eqnarray}
\mathbf{d}\mathbf{e}^{a} & = & \mathbf{e}^{b}\land\boldsymbol{\omega}_{\left(S\right)\;\;b}^{\;a}+\mathbf{S}^{a}\label{Projectively invariant structure equation}
\end{eqnarray}
Comparing the original connection in Eq.(\ref{Original connection})
to $\boldsymbol{\omega}_{\left(S\right)\;b}^{a}$ in Eq.(\ref{Projectively invariant connection}),
\begin{eqnarray*}
\boldsymbol{\omega}_{\;ab}^{\left(S\right)} & = & \boldsymbol{\omega}_{ab}+\frac{1}{2}Q_{\;\;\;bc}^{a}\mathbf{e}^{c}\\
 & = & \frac{1}{2}\left(\boldsymbol{\omega}_{ab}-\boldsymbol{\omega}_{ba}\right)
\end{eqnarray*}
so that $\boldsymbol{\omega}_{\;ab}^{\left(S\right)}$ is simply the
antisymmetric part of the original connection. When the original connection
changes by a projective transformation $\tilde{\boldsymbol{\omega}}_{ab}=\boldsymbol{\omega}_{ab}+\eta_{ab}\boldsymbol{\xi}$
the new connection is unchanged.
\begin{eqnarray*}
\tilde{\boldsymbol{\omega}}_{\;ab}^{\left(S\right)} & = & \tilde{\boldsymbol{\omega}}_{ab}+\frac{1}{2}\tilde{Q}_{\;\;\;bc}^{a}\mathbf{e}^{c}\\
 & = & \boldsymbol{\omega}_{ab}+\eta_{ab}\boldsymbol{\xi}+\frac{1}{2}\left(Q_{abc}-2\eta_{ab}\boldsymbol{\xi}\right)\mathbf{e}^{c}\\
 & = & \boldsymbol{\omega}_{\;ab}^{\left(S\right)}
\end{eqnarray*}

Therefore the $s$-nonmetricity vanishes $Q_{ab}^{\left(S\right)}=-\boldsymbol{\omega}_{ab}^{\left(S\right)}-\boldsymbol{\omega}_{ba}^{\left(S\right)}=0$,
and $\boldsymbol{\omega}_{\left(S\right)ab}=-\boldsymbol{\omega}_{\left(S\right)ba}$
is an $SO\left(p,q\right)$ connection. We have returned to the usual
form of the $\left(p,q\right)$ or Poincarè structure equations, but
now with both manifest projective invariance and local $SO\left(p,q\right)$
invariance.
\begin{eqnarray*}
\mathbf{d}\boldsymbol{\omega}_{\;\;\;b}^{a} & = & \boldsymbol{\omega}_{\left(S\right)\;\;b}^{\;c}\wedge\boldsymbol{\omega}_{\left(S\right)\;\;c}^{\;a}+\boldsymbol{\mathcal{R}}_{\;\;\;b}^{a}\\
\mathbf{d}\mathbf{e}^{a} & = & \mathbf{e}^{b}\wedge\boldsymbol{\omega}_{\left(S\right)\;\;b}^{\;a}+\mathbf{S}^{a}\\
\mathbf{d}\eta_{ab} & = & \eta_{cb}\boldsymbol{\omega}_{\left(S\right)\;\;a}^{\;c}+\eta_{ac}\boldsymbol{\omega}_{\left(S\right)\;\;b}^{\;c}
\end{eqnarray*}

The complete merging of the mixed symmetry subspace of the nonmetricity
tensor $\mathbf{Q}^{a}$ with the torsion is a much stronger relationship
than simple overlap for projective symmetry or dilatations. In the
next Section we show that we may fully rotate $\mathbf{Q}^{a}$ and
$\mathbf{T}^{a}$ into one another without changing the revised structure
equation, Eq.(\ref{Projectively invariant structure equation}). Regardless
of the values of $\mathbf{Q}^{a}$ and $\mathbf{T}^{a}$ separately,
their combination into $\mathbf{S}^{a}$ gives a metric compatible
connection. 

In a straightforward yet nonminimal way, we have generalized the form
of the Cartan equations of the Poincarè group to produce manifest
invariance under both Lorentz and projective transformations while
reproducing the usual form of the Poincarè structure equations.

\subsection{Why does this work?}

The surprising reduction of mixed nonmetricity and torsion into the
single, torsion-like tensor $\mathbf{S}^{a}$ forces us to as whether
there is some deeper symmetry at work. This appears to be the case.
We added Eq.(\ref{New Nonmetricity}) to include the nonmetricity
from the start but it does not have the form of the other structure
equations. It is natural to ask whether the new 2-form $\mathbf{Q}_{a}$
arises from some symmetry.

We wedge with $\frac{1}{2}\mathbf{e}^{b}$ into Eq.(\ref{New Nonmetricity})
to form an equation for $\mathbf{Q}_{a}$ alone.
\begin{eqnarray*}
\frac{1}{2}\mathbf{e}^{b}\wedge\mathbf{d}\eta_{ab} & = & \eta_{cb}\frac{1}{2}\mathbf{e}^{b}\wedge\boldsymbol{\omega}_{\;\;\;a}^{c}+\eta_{ac}\frac{1}{2}\mathbf{e}^{b}\wedge\boldsymbol{\omega}_{\;\;\;b}^{c}+\frac{1}{2}\mathbf{e}^{b}\wedge\mathbf{Q}_{ab}\\
-\mathbf{d}\left(\frac{1}{2}\eta_{ab}\mathbf{e}^{b}\right)+\frac{1}{2}\eta_{ab}\mathbf{d}\mathbf{e}^{b} & = & \frac{1}{2}\eta_{cb}\mathbf{e}^{b}\wedge\boldsymbol{\omega}_{\;\;\;a}^{c}+\frac{1}{2}\eta_{ac}\mathbf{e}^{b}\wedge\boldsymbol{\omega}_{\;\;\;b}^{c}+\mathbf{Q}_{a}
\end{eqnarray*}
Using the Eq.(\ref{New Torsion}) to replace $\mathbf{d}\mathbf{e}^{b}$
and rearanging, this becomes
\begin{eqnarray*}
\mathbf{d}\left(\frac{1}{2}\eta_{ab}\mathbf{e}^{b}\right) & = & \boldsymbol{\omega}_{\;\;\;a}^{c}\wedge\left(\frac{1}{2}\eta_{cb}\mathbf{e}^{b}\right)+\frac{1}{2}\eta_{ab}\mathbf{T}^{b}-\mathbf{Q}_{a}
\end{eqnarray*}
Defining $\mathbf{f}_{a}\equiv\frac{1}{2}\eta_{ab}\mathbf{e}^{b}$
and $\mathbf{U}_{a}\equiv\frac{1}{2}\eta_{ab}\mathbf{T}^{b}+\mathbf{Q}_{a}$
this takes the simple form
\begin{equation}
\mathbf{d}\mathbf{f}_{a}=\boldsymbol{\omega}_{\;\;\;a}^{d}\wedge\mathbf{f}_{d}+\mathbf{U}_{a}\label{Special conformal}
\end{equation}
This is recognizable as the Cartan structure equation of special conformal
transformations, which like dilatations induce a reparameterization
on curves. A moment's reflection reveals the necessity for a transformation
that reparameterizes curves to be related to nonmetricity.

This suggests an alternative decomposition of a general connection.
Rather than separating the connection into compatible, torsion, and
nonmetricity parts, we might consider irreducible representations
and the corresponding vector spaces. Viewed in this way, Young tableau
reduce the $n^{3}$ degrees of freedom of a general connection into
four irreducible subsets. The two mixed symmetry subsets form bases
for the same vector space, so the general connection is spanned by
three vector subspaces.
\begin{itemize}
\item $V_{A}$, the $\frac{1}{6}n\left(n-1\right)\left(n-2\right)$ dimensional
vector space of the totally antisymmetric part of the connection
\item $V_{M}$, a single vector space of dimension $\frac{1}{3}n\left(n^{2}-1\right)$
formed with either set of mixed symmetry components as basis
\item $V_{S}$, the $\frac{1}{6}n\left(n+1\right)\left(n+2\right)$ dimensional
totally symmetric part.
\end{itemize}
This breakdown further suggests repeating the present work with a
top-down approach starting with the auxiliary \cite{WheelerAux} or
biconformal gauging \cite{Wheeler1998,WehnerWheeler,WheelerBic} of
the conformal group. This investigation is in progress \cite{Wheeler2023b}. 

\section{Further invariance of S$^{a}$\label{sec:Further-invariance-of}}

Projective symmetry is not the only invariance of nonminimal compatibility.
Having identified $\mathbf{S}^{a}$ as the sum of two terms it becomes
possible to introduce a larger symmetry. Since the tensors $\mathbf{T}^{a},\mathbf{Q}^{a},\mathbf{S}^{a}$
lie within the vector space of vector-valued 2-forms, we may consider
rotations within that subspace that leave $\mathbf{S}^{a}$ invariant
while mixing $\mathbf{T}^{a}$ and $\mathbf{Q}^{a}$. Clearly, these
will be rotations about $\mathbf{S}^{a}\in\mathcal{A}_{\left[2\right]}^{1}$,
where we define $\mathcal{A}_{\left[2\right]}^{1}$ to be the space
of vector-valued 2-forms.

The vector space of vector-valued 2-forms is large, and any linear
transformation $\phi$ that maps
\begin{eqnarray*}
\phi & : & \mathbf{T}^{a}\rightarrow\tilde{\mathbf{T}}^{a}\\
\phi & : & \mathbf{Q}^{a}=\frac{1}{2}\mathbf{e}^{b}\land\mathbf{Q}_{\;\;\;b}^{a}\rightarrow\tilde{\mathbf{Q}}^{a}\\
\phi & : & \mathbf{S}^{a}\rightarrow\mathbf{S}^{a}
\end{eqnarray*}
also preserves the connection $\boldsymbol{\omega}_{\left(S\right)}^{\;\;ab}$.
Any such transformation $\phi$ therefore preserves the revised Poincarè
structure equations and results in an internal symmetry of general
relativistic spacetimes.

Let $\Sigma:\mathcal{A}_{\left[2\right]}^{1}\leftrightarrow\mathcal{V}$:
be any convenient 1-1 onto linear mapping, $\mathcal{V}=\left\{ V^{A}=\Sigma_{\;\;\;a}^{A\quad bc}V_{\;\;\;bc}^{a},A=1,\ldots,N\right\} $
with norm $g$ induced from the underlying $\left(p,q\right)$ metric
$\eta_{ab}$.
\begin{eqnarray*}
g_{AB} & \equiv & \eta_{ad}\eta_{be}\eta_{cf}\Sigma_{A}^{\;\;\;abc}\Sigma_{B}^{\;\;\;def}
\end{eqnarray*}
The signature $\left(P,Q\right)$ of $g_{AB}$ is determined by $\left(p,q\right)$.
Note that $\dim\mathcal{V}=\frac{1}{2}n^{2}\left(n-1\right)$.

There are three constraints on $\mathbf{S}^{a}$-preserving mappings
$\phi$ of vectors $V^{A}\in\mathcal{V}$. 
\begin{enumerate}
\item $\phi$ must preserve the the $\left(P,Q\right)$ metric $g_{AB}$
induced by the underlying $\left(p,q\right)$ metric.
\[
\left|V^{A}\right|^{2}=g_{AB}V^{A}V^{B}=\frac{1}{2}\eta_{ab}\left(\eta^{ce}\eta^{df}-\eta^{cf}\eta^{de}\right)V_{\;\;\;cd}^{a}V_{\;\;\;ef}^{b}
\]
\item $\phi$ must affect only the mixed symmetry part of $V_{abc}$. The
S-torsion decomposes into independent mixed and totally antisymmetric
parts either of which may or may not enter the field equations. It
is well-known \cite{Datta,HehlDatta,HayashiBreegman,Hehl,HehlvonderHeydeKerlick,HehlNester,HehlNitschvonderHeyde}
that the antisymmetric part of the contorsion $C_{\left[cab\right]}^{\left(S\right)}=-\frac{1}{2}T_{\left[cab\right]}$
is driven by Dirac fields. The transformation $\phi$ must not affect
this part. Couplings to fields such as the spin-$\frac{3}{2}$ Rarita-Schwinger
field--which couples strongly to torsion \cite{Wheeler2023}--may
need modification to be compatible with $\phi$.
\item $\phi$ must preserve $\mathbf{S}^{a}$.
\end{enumerate}

In Appendix \ref{sec:Appendix-B:-Constraints} we find the dependence
of $P$ and $Q$ on $\left(p,q\right)$, and the reduction of the
full $SO\left(P,Q\right)$ which preserves $\mathbf{S}^{a}$ and the
mixed symmetry subspace.

We find that the proper rotation group preserving the metric on $\mathcal{V}$
is 
\begin{eqnarray*}
SO\left(\frac{1}{2}\left(p^{2}\left(p-1\right)+pq\left(3q-1\right)\right),\frac{1}{2}\left(q^{2}\left(q-1\right)+pq\left(3p-1\right)\right)\right)
\end{eqnarray*}
In $4$-dim $\left(3,1\right)$ spacetime this is the split orthogonal
form
\[
SO\left(12,12\right)
\]
This split form $P=Q$ (reminiscent of Kähler, biconformal, and double
field manifolds) occurs if and only if $s=0$ or $n=s^{2}$.

Eliminating totally antisymmetric combinations to affect only the
mixed symmetry subspace reduces this group to

\[
SO\left(\frac{1}{3}p\left(p^{2}+3q^{2}-1\right),\frac{1}{3}q\left(q^{2}+3p^{2}-1\right)\right)
\]
Finally, holding $\mathbf{S}^{a}$ constant reduces the total dimension
$N$ by one. The resulting symmetry group depends on whether $\mathbf{S}^{a}$
is timelike or spacelike. The group is either:
\[
SO\left(\frac{1}{3}p\left(p^{2}+3q^{2}-1\right)-1,\frac{1}{3}q\left(q^{2}+3p^{2}-1\right)\right)
\]
or
\[
SO\left(\frac{1}{3}p\left(p^{2}+3q^{2}-1\right),\frac{1}{3}q\left(q^{2}+3p^{2}-1\right)-1\right)
\]
In $\left(3,1\right)$ spacetime the two possibilities are
\[
SO\left(10,9\right),SO\left(11,8\right)
\]
In either 4-dimensional case the internal symmetry is large enough
to contain the Standard Model, with spacelike $\mathbf{S}^{a}$ leading
directly to the $SO\left(10\right)$ of grand unification. Since the
internal symmetry leaves $\mathbf{S}^{a}$ unchanged, gravity has
decoupled and the Coleman-Mandula theorem is satisfied. 

\subsection{2-form subgroup}

Because $\mathcal{V}$ has the internal structure of $\mathcal{A}_{\left[2\right]}^{1}$,
there are some natural subgroups. For example, it may be useful to
transform the vector and 2-form characters of $\mathcal{V}$ separately.
The vector part of the space is, of course, $n$-dimensional with
signature $\left(p,q\right)$. For the 2-forms we have three cases
with multiplicities
\begin{eqnarray*}
\omega_{a_{1}a_{2}} &  & \frac{1}{2}p\left(p-1\right)\\
\omega_{a_{1}b_{2}} &  & pq\\
\omega_{b_{1}b_{2}} &  & \frac{1}{2}q\left(q-1\right)
\end{eqnarray*}
leading to signature
\begin{eqnarray*}
\left(P_{2},Q_{2}\right) & = & \left(\frac{1}{2}p\left(p-1\right)+\frac{1}{2}q\left(q-1\right),pq\right)
\end{eqnarray*}
For $n>3$ the antisymmetry constriant imposes more than $n$ restrictions,
which therefore cannot be implemented within the vector part alone,
so the $\left(P_{2},Q_{2}\right)$ symmetry will reduce further.

The 4-dim case permits an interesting conjecture. There are 2 positive
norm and 2 negative norm antisymmetry constraints. Three of these
can be imposed on the vector part of the full $A_{\left[2\right]}^{1}$
symmetry. Using the spinor representation $SU\left(2\right)\times SU\left(2\right)$
of the 2-form subgroup $SO\left(P_{2},Q_{2}\right)=SO\left(3,3\right)$
one constraint remains. This must break one of the $SU\left(2\right)$
subgroups, forcing a reduction of an initially left-right symmetric
electroweak model to the actual $SU\left(2\right)\times U\left(1\right)$.

\subsection{Restrictions on Q}

For $\mathbf{S}_{a}$ to be fully general we must have arbitrary $\mathbf{Q}_{a}=\frac{1}{2}\mathbf{e}^{b}\land\mathbf{Q}_{ab}=\frac{1}{2}Q_{a\left[bc\right]}\mathbf{e}^{b}\land\mathbf{e}^{c}$.
We ask whether nonmetricities of the form $Q_{a\left[bc\right]}$
form an invariant vector subspace. The Young tableaux for $\left(\begin{array}{c}
0\\
3
\end{array}\right)$ tensors symmetric on two indices $A_{\left(ab\right)c}$ includes
a totally symmetric part and a mixed symmetry part
\begin{eqnarray*}
n\otimes\left(\frac{n\left(n+1\right)}{2}\right) & = & \frac{1}{6}n\left(n+1\right)\left(n+2\right)\oplus\frac{1}{3}n\left(n^{2}-1\right)
\end{eqnarray*}
This means that the partially symmetrized piece $Q_{a\left(bc\right)}-Q_{\left(abc\right)}$
must be dependent upon $Q_{a\left[bc\right]}$. Checking by adding
and subtracting from a sum of two vectors we find
\begin{eqnarray*}
\frac{1}{3}\left(Q_{b\left[ac\right]}+Q_{c\left[ab\right]}\right) & = & Q_{a\left(bc\right)}-Q_{\left(abc\right)}
\end{eqnarray*}
Therefore, the vector subspaces $\left\{ Q_{a\left[bc\right]}\right\} $
and $\left\{ Q_{\left(abc\right)}\right\} $ are disjoint and span
the full space of nonmetricities. Since the most general form of $S_{abc}$
requires general $\tilde{Q}_{a\left[bc\right]}$ no further reduction
possible. The necessary and sufficient condition we seek is to transform
all nonmetricities $\tilde{Q}_{abc}$ with vanishing totally symmetric
part $Q_{\left(abc\right)}=0$. Neither the totally symmetric part
of $Q_{abc}$ nor the totally antisymmetric part of $T_{abc}$ will
be altered by the internal symmetry.

\section{\label{sec:The-action}The action}

The inclusion of torsion and nonmetricity in the description of gravity
motivates a fresh look at the form of the action.

One motivation for choosing the Einstein-Hilbert action (beyond, of
course, that it works spectacularly) is that when we include the cosmological
constant it is the most general action with field equations of no
more than second order in derivatives of the metric, and linear in
those second derivatives. With torsion and nonmetricity the Einstein-Hilbert
action is no longer the most general action satisfying these conditions.
These tensors depend only on first derivatives of the metric, and
under these criteria may enter quadratically.

In addition to the Einstein-Hilbert term, we consider the most general
action up to quadratic order built from the torsion and non-metricity
as well as the 2- and 3-forms
\begin{eqnarray*}
\mathbf{T} & = & \frac{2}{3}\mathbf{e}^{a}\wedge\mathbf{T}_{a}\\
\mathbf{Q}_{a} & = & \frac{1}{2}\mathbf{e}^{b}\wedge\mathbf{Q}_{ab}
\end{eqnarray*}
It is therefore natural to include the general quadratic combination\footnote{If we allow combinations involving the three traces $S_{a}=S_{\;\;\;ba}^{b},Q_{a}\equiv Q_{\;\;\;ba}^{b},\bar{Q}_{a}\equiv Q_{ab}^{\;\quad b}$
there are two projectively invariant scalars. We may write the most
general projectively invariant action quadratic in the scalars as
\begin{eqnarray*}
S_{additional} & = & \int\left(\mu S_{a}S^{a}+\nu S_{a}^{3}S^{3a}\right)\boldsymbol{\Phi}
\end{eqnarray*}
where $S_{c}^{3}\equiv Q_{c}-n\bar{Q}_{c}$. }
\begin{eqnarray}
S_{Q,T} & = & \int\left(\alpha\mathbf{Q}_{ab}\land\,^{*}\mathbf{Q}^{ab}+\beta\mathbf{Q}_{a}\land\,^{*}\mathbf{Q}^{a}+\mu\mathbf{T}_{a}\land\,^{*}\mathbf{T}^{a}+\nu\mathbf{Q}_{a}\land\,^{*}\mathbf{T}^{a}+\rho\mathbf{T}\land\,^{*}\mathbf{T}\right)\label{S_QT}
\end{eqnarray}
It is interesting to note that there is now coupling between the torsion
and nonmetricity.

The form may now be restricted by requiring projective invariance
of $S_{Q,T}$. Substituting the projective changes of the torsion
and nonmetricity given in Eqs.(\ref{Projective transformation of torsion})
and (\ref{Projective transformation of nonmetricity}), together with
\begin{eqnarray*}
\tilde{\mathbf{T}} & = & \mathbf{T}\\
\tilde{\mathbf{Q}}_{a} & = & \mathbf{Q}_{a}-\eta_{ab}\mathbf{e}^{b}\wedge\boldsymbol{\xi}
\end{eqnarray*}
into Eq.(\ref{S_QT}) and collecting terms, $S_{Q,T}$ is invariant
if and only if $\alpha=0,\beta=-\frac{\nu}{2},\mu=-\frac{\nu}{2}$,
while $\rho$ remains arbitrary. Including these values the most general
action up to linear in second derivatives built from $\mathbf{T}^{a},\mathbf{T},\mathbf{Q}_{ab}$
and $\mathbf{Q}_{a}$ is
\begin{eqnarray}
S_{Q,T} & = & \int\beta\left(\mathbf{T}_{a}-\mathbf{Q}_{a}\right)\land\,^{*}\left(\mathbf{T}^{a}-\mathbf{Q}^{a}\right)+\rho\int\mathbf{T}\land\,^{*}\mathbf{T}\nonumber \\
 & = & \beta\int\mathbf{S}_{a}\land\,^{*}\mathbf{S}^{a}+\rho\int\mathbf{T}\land\,^{*}\mathbf{T}\label{Torsion action}
\end{eqnarray}
Since $\mathbf{T}=\frac{2}{3}\mathbf{e}^{a}\wedge\mathbf{T}_{a}=\frac{2}{3}\mathbf{e}^{a}\wedge\mathbf{S}_{a}\equiv\mathbf{S}$
we may write the full gravitational action as
\begin{eqnarray*}
S_{S}\left[\mathbf{e}^{a},\boldsymbol{\omega}_{\left(S\right)\;\;b}^{\;\;a}\right] & = & \frac{\kappa}{\left(n-2\right)!}\int\boldsymbol{\mathcal{R}}^{ab}\wedge\mathbf{e}^{c}\wedge\ldots\wedge\mathbf{e}^{d}e_{abc\ldots d}\\
 &  & +\beta\int\mathbf{S}_{a}\land\,^{*}\mathbf{S}^{a}+\rho\int\mathbf{S}\land\,^{*}\mathbf{S}
\end{eqnarray*}
The functional $S_{S}$ is Lorentz, projective, and $SO\left(P-1,Q\right)$
(or $SO\left(P,Q-1\right)$) invariant.

Now, in addition to the usual sources we may ask that matter fields
be representations of the internal symmetry, for example, spinor fields
$\psi^{A}$ transforming under the internal rotations, e.g. $Spin\left(P-1,Q\right)$.
Then after gauging we may add
\begin{eqnarray*}
S_{Matter}\left[\mathbf{A}^{B},\psi^{C}\right] & = & \int\alpha g_{AB}\bar{\psi}^{A}\left(i\gamma^{a}D_{a}-m\right)\psi^{B}\boldsymbol{\Phi}+\frac{1}{4}\lambda g_{AB}\mathbf{F}^{A}\wedge^{*}\mathbf{F}^{B}
\end{eqnarray*}
where $\mathbf{F}^{B}=\mathbf{d}\mathbf{A}^{B}+\frac{1}{2}c_{\;\;\;CD}^{B}\mathbf{A}^{C}\wedge\mathbf{A}^{D}$.
Even in 4-dimensions either the $SO\left(10,9\right)$ or $SO\left(11,8\right)$
symmetry is large enough to describe the known interactions.

We vary $S_{S}\left[\mathbf{e}^{a},\boldsymbol{\omega}_{\left(S\right)\;\;b}^{a}\right]$
a la Palatini. We may vary the antisymmetric and symmetric parts of
$S_{S}\left[\mathbf{e}^{a},\boldsymbol{\omega}_{\;\;\;b}^{a},\mathbf{Q}^{a}\right]$
independently, with the symmetric variation is equivalent to varying
$\mathbf{Q}_{ab}$. Alternatively, we may disregard the symmetric
part altogether since the added structure equation
\[
\cancel{\mathbf{d}\eta^{ab}}=-\eta^{cb}\boldsymbol{\omega}_{\left(S\right)\;\;c}^{\;\;a}-\eta^{ac}\boldsymbol{\omega}_{\left(S\right)\;\;c}^{\;\;b}+\cancel{\mathbf{Q}_{\left(S\right)}^{ab}}
\]
now implies metric compatibility while the remaining structure equations
\begin{eqnarray*}
\mathbf{d}\boldsymbol{\omega}_{\left(S\right)\;\;b}^{\;\;a} & = & \boldsymbol{\omega}_{\left(S\right)\;\;b}^{\;\;c}\wedge\boldsymbol{\omega}_{\left(S\right)\;\;c}^{\;\;a}+\boldsymbol{\mathcal{R}}_{\left(C\right)\;\;\;b}^{a}\\
\mathbf{d}\mathbf{e}^{a} & = & \mathbf{e}^{b}\wedge\boldsymbol{\omega}_{\left(S\right)\;\;b}^{\;\;a}+\mathbf{S}^{a}
\end{eqnarray*}
have returned to the Cartan equations of the Poincarè group. 

The internal rotations described here explicitly exclude the totally
antisymmetric part of the torsion, $\mathbf{S}$. However, the totally
antisymmetric part $\mathbf{T}$ of $\mathbf{S}^{a}$ is included
both implicitly in $\mathbf{S}^{a}$ and explicitly in the action
of Eq.(\ref{Torsion action}). This is necessary because the gravitationally
coupled Dirac equation provides a totally antisymmetric source for
torsion \cite{Wheeler2023}.

\section{Conclusions}

We find a large symmetry within Poincarè (or $ISO\left(p,q\right)$)
gauge theory by explicitly allowing both torsion and nonmetricity.
The resulting gravity theory is still a metric compatible Riemann-Cartan
theory of gravity with the internal symmetry decoupled from gravity.
We summarize the steps leading to this conclusion. 

We begin with Poincarè-type gauge theory. Poincarè gauge theory gives
a natural arena for several developments in the theory of gravity.
When the torsion is constrained to zero, it provides a gauge theory
of general relativity, and an arena in which the Palatini variation
is natural. Dropping the constraint on torsion but retaining the Einstein-Hilbert
action gives the well-known ECSK theory of gravity. Even with the
addition of a kinetic term for torsion the theory is consistent with
experiment in certain scenarios. We gave a condensed description of
these geometries in arbitrary dimension $n$ and signature $\left(p,q\right)$.

The next step is an examination of projective transformations. In
affine geometries the connection possesses projective symmetry. This
symmetry arises from reparameterizing autoparallels and preserves
both the autoparallels and the curvature. While Poincarè gauge theory
has both metric and connection, the Palatini variation makes the connection
independent of the metric and we can again consider projective transformations
of the connection. Here the situation is different, for while the
Poincarè geodesics agree with affine autoparallels and the transformations
still preserve both geodesics and the curvature, there are other structures--the
torsion and nonmetricity tensors--which are not projectively invariant. 

This sets up a conflict between the $\left(p,q\right)$ gauge theory
on one hand and our ability to reparameterize geodesics on the other.
We reviewed the well-known Ehlers, Pirani, and Shild resolution to
this dissonance. Extending to an integrable Weyl geometry absorbs
reparameterizations in a manifestly local Lorentz and reparameterization
invariant formalism. We call this the \emph{minimal} modification
of the geometry to achieve the dual invariance.

To achieve manifest local Lorentz and projective invariance we extended
Riemann-Cartan geometry by explicitly including nonmetricity. Within
this geometry we defined a new 2-form tensor given by the difference
of the torsion and the antisymmetric part of the nonmetricity $\mathbf{Q}_{a}\equiv\frac{1}{2}\mathbf{e}^{b}\land\mathbf{Q}_{ab}$.
\[
\mathbf{S}^{a}=\mathbf{T}^{a}-\mathbf{Q}^{a}
\]
This new s-torsion is projectively invariant and has the same vector-valued
2-form symmetry as the original torsion. Surprisingly, we find that
the extended geometry may be written as a metric-compatible geometry
with $\mathbf{S}^{a}$ replacing the torsion. In the new variable
the structure equations reduce to their original Poincarè form even
though the theory is formulated with a fully general connection. This
provides a \emph{nonminimal} means of including reparameterization
invariance within Poincarè gauge theory. The new approach appears
to be the result of using special conformal transformations rather
than dilatations to absorb reparameterizations, but this remains to
be confirmed.

The new torsion field $\mathbf{S}^{a}$ is to be understood as the
physical torsion. Refined Lense-Thirring tests to detect an anomalous
angular momentum, or the sorts of quantum field theory tests described
in \cite{Neville 1980,Shapiro,BelyaevShapiro} can place limits on
the magnitude of torsion. The magnitude of $\mathbf{S}^{a}$ does
not affect the internal symmetry. 

The internal symmetry arises in a way analogous to Wigner's ``little
group'' for a particle. By fixing a particle's 4-momentum and examining
the residual symmetry Wigner identified the $SU\left(2\right)$ rotations
describing spin. Similar considerations apply to higher rank objects.
For example, fixing the Minkowski metric at points of spacetime leaves
local Lorentz transformations as the residual symmetry of spacetime
fields.

Here, the large internal ``little group'' symmetry arises when considering
the space $\mathcal{V}$ of vector-valued 2-forms, which contains
$\mathbf{T}^{a},\mathbf{Q}^{a}$ and $\mathbf{S}^{a}$. In 4-dimensions,
fields of this symmetry have 24 degrees of freedom. Once the field
equations determine the value of the torsion $\mathbf{S}^{a}$, we
may still carry out transformations leaving $\mathbf{S}^{a}$ fixed.

The internal space $\mathcal{V}$ has additional structure, notably
the induced metric $g_{AB}$ built from the spacetime metric, $\eta_{ab}$,
of signature $\left(p,q\right)$. This $\left(p,q\right)$ signature
of $\eta_{ab}$ induces a signature $\left(P,Q\right)$ on $g_{AB}$.
Using the induced metric $g_{AB}$ we checked the number of positive
and negative norm vectors and found the resulting rotation group to
be $SO\left(P,Q\right)$ where 
\begin{eqnarray*}
P & = & \frac{1}{2}\left(p^{2}\left(p-1\right)+pq\left(3q-1\right)\right)\quad\overset{p=3,q=1}{\Longrightarrow}\quad12\\
Q & = & \frac{1}{2}\left(q^{2}\left(q-1\right)+pq\left(3p-1\right)\right)\quad\overset{p=3,q=1}{\Longrightarrow}\quad12
\end{eqnarray*}
This is the $g_{AB}$-preserving symmetry of the full space $\mathcal{V}$.
It reduces to $SO\left(12,12\right)$ for 4-dimensional spacetime.

We applied two reductions of this overall group. First, since the
non-metricity has no totally antisymmetric part we removed the totally
antisymmetric part from $\mathcal{V}$. Second, we considered only
rotations leaving the effective torsion $\mathbf{S}^{a}$ invariant.
The final internal symmetry is then $SO\left(P',Q'\right)$ with
\begin{eqnarray*}
P' & = & \frac{1}{3}p\left(p^{2}+3q^{2}-1\right)-1\quad\overset{p=3,q=1}{\Longrightarrow}\quad10\\
Q' & = & \frac{1}{3}q\left(q^{2}+3p^{2}-1\right)\quad\quad\quad\overset{p=3,q=1}{\Longrightarrow}\quad9
\end{eqnarray*}
or
\begin{eqnarray*}
P' & = & \frac{1}{3}p\left(p^{2}+3q^{2}-1\right)\quad\quad\quad\overset{p=3,q=1}{\Longrightarrow}\quad11\\
Q' & = & \frac{1}{3}q\left(q^{2}+3p^{2}-1\right)-1\quad\overset{p=3,q=1}{\Longrightarrow}\quad8
\end{eqnarray*}
Either of these groups contains symmetry sufficient for the standard
model.

The gravitational effects of these models depend on the torsion $\mathbf{S}^{a}$
and the corresponding curvature. Since $\mathbf{S}^{a}$ is always
orthogonal to the rotations, hence invariant, the $SO\left(P',Q'\right)$
symmetry decouples from gravity, giving an internal symmetry in agreement
with the Coleman-Mandula theorem. This internal symmetry depends only
on the presence of $\mathbf{S}^{a}$, not on its magnitude or the
strength of its couplings to other fields.

In 4-dimensional spacetime $SO\left(12,12\right)$ preserves the induced
metric, and the $\mathbf{S}^{a}$-preserving internal symmetry group
is $SO\left(10,9\right),Spin\left(10,9\right),SO\left(11,8\right)$
or $Spin\left(11,8\right)$. This opens the possibility of a fully
unified theory without the need for higher dimensions.

The most general action up to linearity in second derivatives of the
solder form contains six terms. These include combinations quadratic
in torsion and nonmetricity as well as torsion-nonmetricity couplings,
in addition to the Einstein-Hilbert action. Imposing projective invariance
reduces this to three terms dependent on S and curvature only.

\section*{Appendix A: Formal development of SO(p,q) geometry\label{sec:Appendix-A:-Formal}}

We develop the arena for local $SO\left(p,q\right)$ symmetric physical
models based on the unrestricted Cartan gauge theory of the Lie group
$\mathcal{P}=ISO\left(p,q\right)$. Additional discussion may be found
in \cite{Wheeler2023}.

Starting with the Maurer-Cartan equations of $\mathcal{P}$
\begin{eqnarray*}
\mathbf{d}\tilde{\boldsymbol{\omega}}_{\;\;\;b}^{a} & = & \tilde{\boldsymbol{\omega}}_{\;\;\;b}^{c}\wedge\tilde{\boldsymbol{\omega}}_{\;\;\;c}^{a}\\
\mathbf{d}\tilde{\mathbf{e}}^{a} & = & \tilde{\mathbf{e}}^{b}\wedge\tilde{\boldsymbol{\omega}}_{\;\;\;b}^{a}
\end{eqnarray*}
we take the quotient by the Lie subgroup $\mathcal{L}=SO\left(p,q\right)$
. The projection from cosets $\mathcal{L}_{g}$ to the homogeneous
quotient manifold $\mathcal{M}^{n}=\mathcal{P}/\mathcal{L}$ allows
us to develop a principal fiber bundle with $SO\left(p,q\right)$
symmetry. By modifying the solder form and the spin connection 1-forms
$\left(\tilde{\mathbf{e}}^{b},\tilde{\boldsymbol{\omega}}_{\;\;\;b}^{a}\right)\rightarrow\left(\mathbf{e}^{b},\boldsymbol{\omega}_{\;\;\;b}^{a}\right)$
we introduce a $\mathcal{P}$-covariant curvature 2-form with two
$\mathcal{L}$-covariant components: the \emph{curvature} $\boldsymbol{\mathcal{R}}_{\;\;\;b}^{a}$
and the \emph{torsion} $\mathbf{T}^{a}$
\begin{eqnarray}
\mathbf{d}\boldsymbol{\omega}_{\;\;\;b}^{a} & = & \boldsymbol{\omega}_{\;\;\;b}^{c}\wedge\boldsymbol{\omega}_{\;\;\;c}^{a}+\boldsymbol{\mathcal{R}}_{\;\;\;b}^{a}\nonumber \\
\mathbf{d}\mathbf{e}^{a} & = & \mathbf{e}^{b}\wedge\boldsymbol{\omega}_{\;\;\;b}^{a}+\mathbf{T}^{a}\label{Cartan equations}
\end{eqnarray}
We require the $\boldsymbol{\mathcal{R}}_{\;\;\;b}^{a}$ and $\mathbf{T}^{a}$
to be horizontal,
\begin{eqnarray*}
\boldsymbol{\mathcal{R}}_{\;\;\;b}^{a} & = & \frac{1}{2}\mathcal{R}_{\;\;\;bcd}^{a}\mathbf{e}^{c}\land\mathbf{e}^{d}\\
\mathbf{T}^{a} & = & \frac{1}{2}T_{\;\;\;bc}^{a}\mathbf{e}^{b}\land\mathbf{e}^{c}
\end{eqnarray*}
thereby preserving the bundle structure by making integrals of the
connection independent of lifting. Integrability of the Cartan equations
Eqs.(\ref{Cartan equations}) is insured by $\mathbf{d}^{2}\boldsymbol{\omega}_{\;\;\;b}^{a}\equiv0$
and $\mathbf{d}^{2}\mathbf{e}^{a}\equiv0$, which lead to the Bianchi
identities,
\begin{eqnarray}
\boldsymbol{\mathcal{D}}\mathbf{T}^{a} & = & \mathbf{e}^{b}\land\boldsymbol{\mathcal{R}}_{\;\;\;b}^{a}\nonumber \\
\boldsymbol{\mathcal{D}}\boldsymbol{\mathcal{R}}_{\;\;\;b}^{a} & = & 0\label{Bianchis}
\end{eqnarray}
The covariant exterior derivatives are given by
\begin{eqnarray*}
\boldsymbol{\mathcal{D}}\boldsymbol{\mathcal{R}}_{\;\;\;b}^{a} & = & \mathbf{d}\boldsymbol{\mathcal{R}}_{\;\;\;b}^{a}+\boldsymbol{\mathcal{R}}_{\;\;\;b}^{c}\wedge\boldsymbol{\omega}_{\;\;\;c}^{a}-\boldsymbol{\omega}_{\;\;\;b}^{c}\wedge\boldsymbol{\mathcal{R}}_{\;\;\;c}^{a}\\
\boldsymbol{\mathcal{D}}\mathbf{T}^{a} & = & \mathbf{d}\mathbf{T}^{a}+\mathbf{T}^{b}\wedge\boldsymbol{\omega}_{\;\;\;b}^{a}
\end{eqnarray*}
The frame field $\mathbf{e}^{a}$ is taken $\left(p,q\right)$-orthonormal
$\left\langle \mathbf{e}^{a},\mathbf{e}^{b}\right\rangle =\eta^{ab}=diag\left(1,\ldots,1,-1,\ldots,-1\right)$
with the connection assumed to be metric compatible
\begin{eqnarray*}
\mathbf{d}\eta^{ab}+\eta^{cb}\boldsymbol{\omega}_{\;\;\;c}^{a}+\eta^{ac}\boldsymbol{\omega}_{\;\;\;c}^{b} & = & 0
\end{eqnarray*}
Since $\mathbf{d}\eta^{ab}=0$, the spin connection is antisymmetric,
$\boldsymbol{\omega}_{ab}=-\boldsymbol{\omega}_{ba}$.

The equations above describe Riemann-Cartan geometry in the Cartan
formalism. Note that the Riemann-Cartan curvature, $\boldsymbol{\mathcal{R}}_{\;\;\;b}^{a}$,
differs from the Riemann curvature $\mathbf{R}_{\;\;\;b}^{a}$ by
terms dependent on the torsion. 

When the torsion vanishes, $\mathbf{T}^{a}=0$, the Riemann-Cartan
curvature $\boldsymbol{\mathcal{R}}_{\;\;\;b}^{a}$ reduces to the
Riemann curvature $\mathbf{R}_{\;\;\;b}^{a}$ and Eqs.(\ref{Cartan equations})
exactly reproduce the expressions for the connection and curvature
of a general Riemannian geometry. At the same time, Eqs.(\ref{Bianchis})
reduce to the usual first and second Bianchi identities. We may constrain
$\mathbf{T}^{a}=0$ in the Cartan equations of Riemann-Cartan geometry,
reducing the structure equations to those of Riemannian geometry with
its known consistency. 

These results are geometric; a physical model follows when we posit
an action functional. The action may depend on the bundle tensors
$\mathbf{e}^{b},\mathbf{T}^{a},\boldsymbol{\mathcal{R}}_{\;\;\;b}^{a}$
and the invariant tensors $\eta_{ab}$ and $e_{ab\ldots d}$. To this
we may add source functionals built from any field representations
of the fiber symmetry group $\mathcal{L}$, including scalars, spinors,
vector fields, etc.

Constraining the torsion zero, specifying the Einstein-Hilbert form
of action, and varying only the solder form, the $q=1$ theory describes
general relativity as a gauge theory in $n$-dimensions. We cannot
vary the metric and connection independently because this can introduce
nonzero sources for torsion, making the $\mathbf{T}^{a}=0$ constraint
inconsistent.

Dropping the torsion constraint while retaining the Einstein-Hilbert
action gives the Einstein-Cartan-Sciama-Kibble (ECSK) theory of gravity
in Riemann-Cartan geometry. The torsion is found to depend on the
spin tensor, given by the connection variation of the source $\sigma$$_{\;\;\;ab}^{\mu}$$=\frac{\delta L}{\delta\omega_{\;\;\mu}^{ab}}$.
Without modifying the action to include dynamical torsion, the resulting
torsion survives only within matter.

\subsection*{Solving for the connection}

\subsubsection*{Contorsion}

The Cartan structure equations (\ref{Cartan equations}), allow us
to derive an explicit form for the connection and reduced form for
the curvature. Starting from the equation for the torsion
\begin{eqnarray*}
\mathbf{d}\mathbf{e}^{a} & = & \mathbf{e}^{b}\wedge\boldsymbol{\omega}_{\;\;\;b}^{a}+\mathbf{T}^{a}
\end{eqnarray*}
write the spin connection as the sum of two terms
\[
\boldsymbol{\omega}_{\;\;\;b}^{a}=\boldsymbol{\alpha}_{\;\;\;b}^{a}+\boldsymbol{\beta}_{\;\;\;b}^{a}
\]
where $\boldsymbol{\alpha}_{ab}=-\boldsymbol{\alpha}_{ba}$ is the
torsion-free connection, $\mathbf{d}\mathbf{e}^{a}=\mathbf{e}^{b}\land\boldsymbol{\alpha}_{\;\;\;b}^{a}$
and $\boldsymbol{\beta}_{ab}=-\boldsymbol{\beta}_{ba}$. Then $\boldsymbol{\beta}_{\;\;\;b}^{a}$
must satisfy
\[
0=\mathbf{e}^{b}\land\boldsymbol{\beta}_{\;\;\;b}^{a}+\mathbf{T}^{a}
\]
To solve this the 1-form $\boldsymbol{\beta}_{ab}$ will be linear
in the torsion and antisymmetric. These conditions dictate the ansatz
\begin{eqnarray*}
\boldsymbol{\beta}_{ab} & = & \left(aT_{cab}+b\left(T_{acb}-T_{bca}\right)\right)\mathbf{e}^{c}
\end{eqnarray*}
for some constants $a,b$. Substitution quickly leads to $a=b=\frac{1}{2}$,
and the spin connection is
\begin{equation}
\boldsymbol{\omega}_{\;\;\;b}^{a}=\boldsymbol{\alpha}_{\;\;\;b}^{a}+\mathbf{C}_{\;\;\;b}^{a}\label{Spin connection decomposition}
\end{equation}
where $\mathbf{C}_{\;\;\;b}^{a}$ is the \emph{contorsion},
\[
\mathbf{C}_{\;\;\;b}^{a}=\frac{1}{2}\left(T_{c\;\quad b}^{\;\;\;a}+T_{b\;\;\quad c}^{\;\;\;a}-T_{\;\;\;bc}^{a}\right)\mathbf{e}^{c}
\]
Contorsion transforms tensorially so this form is unique. We may recover
the torsion by wedging and contracting with $\mathbf{e}^{b}$.
\begin{eqnarray*}
\mathbf{C}_{\;\;\;b}^{a}\wedge\mathbf{e}^{b} & = & \mathbf{T}^{a}
\end{eqnarray*}

The torsion now enters the curvature through the connection. Expanding
the Cartan-Riemann curvature $\boldsymbol{\mathcal{R}}_{\;\;\;b}^{a}$
using Eq.(\ref{Spin connection decomposition}) and identifying the
$\boldsymbol{\alpha}$-covariant derivative, $\mathbf{D}\mathbf{C}_{\;\;\;b}^{a}=\mathbf{d}\mathbf{C}_{\;\;\;b}^{a}-\mathbf{C}_{\;\;\;b}^{c}\land\boldsymbol{\alpha}_{\;\;\;c}^{a}-\boldsymbol{\alpha}_{\;\;\;b}^{c}\land\mathbf{C}_{\;\;\;c}^{a}$
leads to
\begin{equation}
\boldsymbol{\mathcal{R}}_{\;\;\;b}^{a}=\mathbf{R}_{\;\;\;b}^{a}+\mathbf{D}\mathbf{C}_{\;\;\;b}^{a}-\mathbf{C}_{\;\;\;b}^{c}\land\mathbf{C}_{\;\;\;c}^{a}\label{Curvature decomposition}
\end{equation}
This is the Riemann-Cartan curvature expressed in terms of the Riemann
curvature and the contorsion. If we contract with $\mathbf{e}^{b}$
we recover the Bianchi identity. This happens because our solution
for the connection automatically satisfies the integrability condition
for the connection.

\subsubsection*{Bianchi identities}

Given Eq.(\ref{Curvature decomposition}) for the Riemann-Cartan curvature,
we may also expand the generalized Bianchi identities (\ref{Bianchis}).
The first Bianchi becomes
\begin{eqnarray*}
\mathbf{d}\mathbf{T}^{a}+\mathbf{T}^{b}\wedge\left(\boldsymbol{\alpha}_{\;\;\;b}^{a}+\mathbf{C}_{\;\;\;b}^{a}\right) & = & \mathbf{e}^{b}\wedge\mathbf{R}_{\;\;\;b}^{a}+\mathbf{e}^{b}\wedge\mathbf{D}\mathbf{C}_{\;\;\;b}^{a}-\mathbf{e}^{b}\wedge\mathbf{C}_{\;\;\;b}^{c}\land\mathbf{C}_{\;\;\;c}^{a}
\end{eqnarray*}
Using $\mathbf{C}_{\;\;\;b}^{c}\wedge\mathbf{e}^{b}=\mathbf{T}^{c}$
and $\mathbf{D}\mathbf{e}^{a}=0$ the torsion terms cancel and we
may write $\mathbf{e}^{b}\wedge\mathbf{D}\mathbf{C}_{\;\;\;b}^{a}=\mathbf{D}\left(\mathbf{C}_{\;\;\;b}^{a}\wedge\mathbf{e}^{b}\right)=\mathbf{D}\mathbf{T}^{a}$.
The Riemannian Bianchi $\mathbf{e}^{b}\wedge\mathbf{R}_{\;\;\;b}^{a}=0$
follows immediately. Similarly, expanding the derivative in the second
Bianchi gives
\[
0=\mathbf{D}\boldsymbol{\mathcal{R}}_{\;\;\;b}^{a}+\boldsymbol{\mathcal{R}}_{\;\;\;b}^{c}\wedge\mathbf{C}_{\;\;\;c}^{a}-\mathbf{C}_{\;\;\;b}^{c}\wedge\boldsymbol{\mathcal{R}}_{\;\;\;c}^{a}
\]
and replacing $\boldsymbol{\mathcal{R}}_{\;\;\;b}^{a}=\mathbf{R}_{\;\;\;b}^{a}+\mathbf{D}\mathbf{C}_{\;\;\;b}^{a}-\mathbf{C}_{\;\;\;b}^{c}\land\mathbf{C}_{\;\;\;c}^{a}$
throughout then using $\mathbf{C}_{\;\;\;b}^{c}\wedge\mathbf{e}^{b}=\mathbf{T}^{c}$
and the Ricci identity $\mathbf{D}^{2}\mathbf{C}_{\;\;\;b}^{a}=\mathbf{C}_{\;\;\;b}^{c}\land\mathbf{R}_{\;\;\;c}^{a}-\mathbf{C}_{\;\;\;c}^{a}\land\mathbf{R}_{\;\;\;b}^{c}$
lead to cancellations down to the second Riemannian Bianchi identity
\begin{eqnarray*}
\mathbf{D}\mathbf{R}_{\;\;\;b}^{a} & = & 0
\end{eqnarray*}
Therefore, the Cartan-Riemann Bianchi identities hold if and only
if the Riemann Bianchi identities hold.

Because the curvature is a 2-form, and the spin connection is antisymmetric,
$\mathcal{R}_{abcd}=\mathcal{R}_{ab\left[cd\right]}=\mathcal{R}_{\left[ab\right]cd}$
and there is still only one independent contraction of the curvature.
The first Bianchi identity then shows that the curvature tensor $\boldsymbol{\mathcal{R}}_{\;\;\;b}^{a}$
has nonvanishing triply antisymmetric part. Expanding both sides of
the first Bianchi identity in components and antisymmetrizing we take
a single contraction to show an antisymmetric part
\begin{eqnarray}
\mathcal{R}_{ba}-\mathcal{R}_{ab} & = & \mathcal{D}_{c}\mathscr{T}_{\;\;\;ab}^{c}\label{Antisymmetric Ricci-1}
\end{eqnarray}
where $\mathscr{T}_{\;\;\;ab}^{c}\equiv T_{\;\;\;ab}^{c}-\delta_{a}^{c}T_{\;\;\;db}^{d}+\delta_{b}^{c}T_{\;\;\;da}^{d}$,
so the Ricci tensor of the Cartan-Riemann curvature acquires an antisymmetric
part dependent on derivatives of the torsion.

\section*{Appendix B: Constraints on rotations\label{sec:Appendix-B:-Constraints}}

We impose three conditions on linear mappings $\phi$ on vectors $V^{A}\in\mathcal{V}$. 
\begin{enumerate}
\item $\phi$ must preserve the the $\left(P,Q\right)$ metric $g_{AB}$
induced by the underlying $\left(p,q\right)$ metric.
\[
\left|V^{A}\right|^{2}=g_{AB}V^{A}V^{B}=\frac{1}{2}\eta_{ab}\left(\eta^{ce}\eta^{df}-\eta^{cf}\eta^{de}\right)V_{\;\;\;cd}^{a}V_{\;\;\;ef}^{b}
\]
\item $\phi$ must affect only the mixed symmetry part of $V_{abc}$.
\item $\phi$ must preserve $\mathbf{S}^{a}$.
\end{enumerate}
We look at each condition in turn. 

\subsection*{Metric}

We begin by finding the metric signature $\left(P,Q\right)$. Let
the spacelike and timelike components of any vector $V^{a}$ be separated
as $V^{a}=\left(V^{a_{i}},V^{b_{j}}\right)$ with $a_{i};i=1,\ldots,p$
and $b_{j};j=1,\ldots,q$. Then the norm of $V_{abc}$ is
\begin{eqnarray*}
\left|V^{A}\right|^{2} & = & \sum_{a_{1};a_{2}<a_{3}}\left(V_{a_{1}a_{2}a_{3}}\right)^{2}-\sum_{b_{1};a_{1}<a_{2}}\left(V_{b_{1}a_{1}a_{2}}\right)^{2}-\sum_{a_{1},b_{1},a_{2}}\left(V_{a_{1}b_{1}a_{2}}\right)^{2}\\
 &  & +\sum_{a_{1},b_{1},b_{2}}\left(V_{b_{1}b_{2}a_{1}}\right)^{2}+\sum_{a_{1},b_{1}<b_{2}}\left(V_{a_{1}b_{1}b_{2}}\right)^{2}-\sum_{b_{1},b_{2}<b_{3}}\left(V_{b_{1}b_{2}b_{3}}\right)^{2}
\end{eqnarray*}
The multiplicities of the positive sums are, respectively: $\frac{1}{2}p^{2}\left(p-1\right),q^{2}p,\frac{1}{2}pq\left(q-1\right)$.
For the negative sums multiplicities are: $\frac{1}{2}qp\left(p-1\right),p^{2}q,\frac{1}{2}q^{2}\left(q-1\right)$.
Combining
\begin{eqnarray*}
P & = & \frac{1}{2}\left(p^{2}\left(p-1\right)+pq\left(3q-1\right)\right)\\
Q & = & \frac{1}{2}\left(q^{2}\left(q-1\right)+pq\left(3p-1\right)\right)
\end{eqnarray*}
with $P+Q=N=\frac{1}{2}n^{2}\left(n-1\right)$ and $S=P-Q=\frac{1}{2}s\left(s^{2}-n\right)$.

The proper rotation group preserving the metric on $\mathcal{V}$
is therefore 
\begin{eqnarray*}
SO\left(\frac{1}{2}\left(p^{2}\left(p-1\right)+pq\left(3q-1\right)\right),\frac{1}{2}\left(q^{2}\left(q-1\right)+pq\left(3p-1\right)\right)\right)
\end{eqnarray*}
For example, in $4$-dim $\left(3,1\right)$ spacetime this is the
split orthogonal form
\[
SO\left(12,12\right)
\]
This split form $P=Q$ (reminiscent of Kähler, biconformal, and double
field manifolds) occurs if and only if $s=0$ or $n=s^{2}$ is a perfect
square.

\subsection*{Antisymmetry constraint}

The antisymmetry constraints fall into four types, each of definite
causality type in the induced norm:
\begin{eqnarray*}
X_{1}=V_{a_{1}a_{2}a_{3}}+V_{a_{3}a_{1}a_{2}}+V_{a_{2}a_{3}a_{1}} & = & 0\quad(spacelike)\\
X_{2}=V_{b_{1}a_{2}a_{3}}+V_{a_{3}b_{1}a_{2}}+V_{a_{2}a_{3}b_{1}} & = & 0\quad(timelike)\\
X_{3}=V_{b_{1}b_{2}a_{3}}+V_{a_{3}b_{1}b_{2}}+V_{b_{2}a_{3}b_{1}} & = & 0\quad(spacelike)\\
X_{4}=V_{b_{1}b_{2}b_{3}}+V_{b_{3}b_{1}b_{2}}+V_{b_{2}b_{3}b_{1}} & = & 0\quad(timelike)
\end{eqnarray*}
It is straightforward to count the multiplicities, bearing in mind
that the three indices must all differ:
\begin{eqnarray*}
X_{1} &  & \frac{p\left(p-1\right)\left(p-2\right)}{3!}\\
X_{2} &  & \frac{qp\left(p-1\right)}{2!}\\
X_{3} &  & \frac{pq\left(q-1\right)}{2!}\\
X_{4} &  & \frac{q\left(q-1\right)\left(q-2\right)}{3!}
\end{eqnarray*}
and we check that these sum to the required $\frac{1}{3!}n\left(n-1\right)\left(n-2\right)$
components of $V_{\left[abc\right]}$.

Setting each set of components $X_{i}=0$ reduces $P$ and $Q$ for
the symmetry to
\begin{eqnarray*}
P & = & \frac{1}{2}\left(p^{3}+3q^{2}p-p^{2}-pq\right)-\frac{p\left(p-1\right)\left(p-2\right)}{3!}-\frac{pq\left(q-1\right)}{2!}\\
 & = & \frac{1}{3}p\left(p^{2}+3q^{2}-1\right)\\
Q & = & \frac{1}{2}\left(-qp+3p^{2}q+q^{3}-q^{2}\right)-\frac{qp\left(p-1\right)}{2!}-\frac{q\left(q-1\right)\left(q-2\right)}{3!}\\
 & = & \frac{1}{3}\left(q^{3}+3p^{2}q-q\right)
\end{eqnarray*}
The reduced group is now
\[
SO\left(\frac{1}{3}p\left(p^{2}+3q^{2}-1\right),\frac{1}{3}q\left(q^{2}+3p^{2}-1\right)\right)
\]
The representation is split if and only if $p-q=0,\pm1$. In $\left(3,1\right)$
spacetime this reduces the internal symmetry from $SO\left(12,12\right)$
to $SO\left(11,9\right)$.

\subsection*{Constant $\mathbf{S}{}^{a}$}

Holding $\mathbf{S}^{a}$ constant reduces the total dimension $N$
by one. The resulting symmetry group depends on whether $\mathbf{S}^{a}$
is timelike or spacelike. The group is either:
\[
SO\left(\frac{1}{3}p\left(p^{2}+3q^{2}-1\right)-1,\frac{1}{3}q\left(q^{2}+3p^{2}-1\right)\right)
\]
or
\[
SO\left(\frac{1}{3}p\left(p^{2}+3q^{2}-1\right),\frac{1}{3}q\left(q^{2}+3p^{2}-1\right)-1\right)
\]
In $\left(3,1\right)$ spacetime the two possibilities are
\[
SO\left(10,9\right),SO\left(11,8\right)
\]
In either case the internal symmetry is large enough to contain the
Standard Model, with spacelike $\mathbf{S}^{a}$ leading directly
to the $SO\left(10\right)$ of grand unification.
\end{document}